\documentclass[aps,pre,reprint,longbibliography,showpacs,superscriptaddress,groupaddress,showkeys,floatfix,byrevtex,]{revtex4-1}
\usepackage{graphicx}
\usepackage{dcolumn}
\usepackage{bm}
\usepackage[utf8x]{inputenc}
\usepackage{srctex}
\usepackage{amsmath}
\usepackage{amsfonts}
\usepackage{amssymb}
\usepackage{graphicx}
\usepackage{overpic}
\usepackage{srctex}
\usepackage{xcolor}
\usepackage{subfigure}
\usepackage{cancel}
\usepackage[normalem]{ulem}

\begin{document}

\title{Stationary superstatistics distributions of trapped run-and-tumble particles}

\author{Francisco J. Sevilla}
\email[]{fjsevilla@fisica.unam.mx}
\thanks{}
\affiliation{Instituto de F\'isica, Universidad Nacional Aut\'onoma de M\'exico, Apdo.\ Postal 20-364, 01000, Ciudad de M\'exico, Mexico}

\author{Alejandro \surname{V. Arzola}}
\affiliation{Instituto de F\'isica, Universidad Nacional Aut\'onoma de M\'exico, Apdo.\ Postal 20-364, 01000, Ciudad de M\'exico, Mexico}

\author{Enrique \surname{Puga Cital}}
\affiliation{Instituto de F\'isica, Universidad Nacional Aut\'onoma de M\'exico, Apdo.\ Postal 20-364, 01000, Ciudad de M\'exico, Mexico}

\date{Today}

\begin{abstract}
We present an analysis of the stationary distributions of run-and-tumble particles trapped in external potentials in terms of a thermophoretic potential, that 
emerges when trapped active motion is mapped to trapped passive Brownian motion in a fictitious inhomogeneous thermal bath. We 
elaborate on the meaning of the non-Boltzmann-Gibbs 
stationary distributions that emerge as a consequence of the persistent motion of active particles. These stationary distributions are interpreted as a 
class of distributions in nonequilibrium statistical mechanics known as superstatistics. Our analysis provides an original insight on the link between the intrinsic 
nonequilibrium nature of active motion and the well-known concept of local equilibrium used in nonequilibrium statistical mechanics, and contributes to the understanding of 
the validity of the concept of effective temperature. Particular cases of interest, regarding specific trapping potentials used in other theoretical or experimental studies, 
are discussed. We point out as an unprecedented effect, the emergence of new modes of the stationary distribution as a consequence of the coupling of persistent motion in a 
trapping potential that 
varies highly enough with position.
\end{abstract}


\maketitle

\section{Introduction}

Self-propelled or active particles are open systems driven out of thermal equilibrium by complex internal mechanisms that locally convert energy from the environment into 
active motion \cite{RamaswamyAnnRevCondMattPhys2010,MarchettiRMP2013,BechingerRMP2016}. A variety of microorganisms suspended in temperate aqueous environments use internal 
motors that allow them to move, undulate, or rotate flagella or cilia at small Reynolds numbers to self-propel leading to a variety of patterns of motion that are of 
great interest to statistical physicists. On the other hand, different phoretic mechanisms, such as thermophoresis and diffusiophoresis, have been ingeniously devised to 
endow passive Brownian particles with active motion.

Two main features characterize active motion: first, the tendency of particles to move at a characteristic speed as a consequence of self-propulsion, and the other, the 
\emph{persistence} or tendency to maintain the direction of motion for long enough intervals of time. Though it is clear that active motion corresponds to the class of 
intrinsically out-of-equilibrium phenomena, it is often difficult to give a measure for the departure from equilibrium \cite{FodorPRL2016}. Notwithstanding this, today there 
has been great effort made to give a thermodynamic description of active matter \cite{SolonPRL2015,TakatoriCurrentOpinion2016}. On this course, the concept of effective 
temperature $T_{\text{eff}}$ is valuable and has been explored theoretically and experimentally in a variety of systems in nonequilibrium 
situations
\cite{OukrisNphys2010,ColombaniPRL2011,NobrePRE2012,DieterichNaturePhys2015,LoiPRE2008,TailleurEPL2009,PalacciPRL2010,EnculescuPRL2011,BenIsaacPRL2011,LoiSoftMatterM2011,
SzamelPRE2014,LevisEPL2015}, particularly in the dilute regime \cite{FilyPRL2012}. More recently, it has been shown that the effective temperature in a two-component active 
Janus particles can be considered a control parameter (in the sense of a thermodynamics variable) for the observed kinetics and phase behavior \cite{HanPNAS2017}. 

The physical intuition behind the concept of effective temperature relies on the fulfillment of the fluctuation-dissipation relation. The trajectories of 
active particles, obtained from experiments \cite{PalacciPRL2010,DieterichNaturePhys2015}, numerical simulations \cite{LoiPRE2008,LoiSoftMatterM2011}, or analytical results 
\cite{TailleurEPL2009,EnculescuPRL2011}, have allowed to conclude that an effective temperature emerges from the internal fluctuations of the active motion, which are of no 
thermal origin and that effectively emulate thermal fluctuations. Indeed, $T_{\text{eff}}$ can be determined by using a tracer as thermometer that probes a nonequilibrium 
complex media through a diffusion process. Experimental realizations of this idea are numerous, for instance the motion of a bead in a bath of bacteria can exhibit effective 
temperatures as large as two or three orders of magnitude of the room temperature \cite{WuPRL2000}.

Active particles freely moving at constant speed $v$ with characteristic persistence time $\alpha^{-1}$ exhibit normal diffusion in the long-time regime, the diffusion 
constant being $D_{\text{eff}}=v^{2}/\alpha$. In this regime the motion of an active particle can \emph{equivalently} be thought of as the motion of a passive one in a 
\emph{fictitious} hotter environment with effective temperature $T_{0}=D_{\text{eff}}/k_{B}\mu$, with $\mu$ and $k_{B}$ being the mobility and the Boltzmann constant, 
respectively \cite{SchnitzerPRE1993,TailleurEPL2009}. For instance, dilute suspensions of self-propelled particles in sedimentation processes can be considered as 
passive Brownian particles in a hotter source of heat with an effective temperature that scales linearly with the persistence time \cite{PalacciPRL2010}.

In systems of confined active particles, either by impenetrable walls or by external potentials 
\cite{ErdmannEPJB2000,TailleurEPL2009,FilyPRL2012,TakatoriNatcomms2016,BechingerRMP2016,SartoriPRE2018}, the situation is quite different. Indeed, the effects of persistence 
are conspicuously revealed in the long time regime if the persistence length is larger than the characteristic length of confinement. Notoriously, the stationary 
distribution of the particle positions shows an accumulation of particles around the boundaries of confinement (see, for instance, Ref.~\cite{MiramontesPlos2014} for 
trajectories of worker termites in a circular arena, Ref. \cite{BricardNatureComm2015} for confined colloidal rollers in a circular disk, or Ref.~\cite{TakatoriNatcomms2016} 
for active Brownian particles confined in an acoustic trap), corresponding evidently to non-Boltzmann-Gibbs distributions \cite{TailleurPRL2008,Solon2015EPJST2015}. Such 
non-Boltzmann-Gibbs distributions have also been observed experimentally in optically trapped passive Brownian particles coupled to a bath of active ones 
\cite{ArgunPRE0216}, in theoretically described tracer particles diffusing in an elastic active gel \cite{Ben-IsaacPRE2015}, and in a model of active glasses 
\cite{NandiPNAS2018}, which manifestly exhibits the intrinsic nonequilibrium nature of active baths.

Is it possible to describe the non-Boltzmann-Gibbs stationary distributions of noninteracting active particles trapped in an 
external potential, in terms of passives ones in a fictitious environment? The answer is in the positive if the concept of effective temperature (homogeneous) is extended to a
nonequilibrium 
(inhomogeneous) environment. To be more explicit, we found that for spatially independent parameters that characterize the active 
motion under consideration (a situation that can be thought as an effective homogeneous medium  of ``nutrients''), such a mapping can be identified in terms 
of an effective inhomogeneous temperature $\mathcal{T}(x)$. The mapping allows us to make an analysis based on the stochastic thermodynamics, but, more importantly, it 
provides a powerful tool to pinpoint the nonequilibrium nature of active motion. In particular, it allows 
us to interpret these distributions as a class of the distributions that appear in nonequilibrium statistical mechanics known as 
\emph{superstatistics} \cite{AbePRE2007,BeckPRE2005}.

Our theoretical analysis is based on the active motion described by the \emph{run-and-tumble} dynamics, a mathematical framework used by several authors 
 which corresponds to a generalization of the telegrapher process \cite[and references therein]{SchnitzerPRE1993, TailleurPRL2008}. Such a framework is presented in some detail, for 
the sake of completeness, in Sec. \ref{section2}. In this same section the mapping between trapped active motion and trapped passive one in an 
inhomogeneous environment is devised and some results  are formulated. In Sec. \ref{section3}, the consequences of the devised mapping are 
analyzed within the framework of the stochastic thermodynamics for the case of one-dimensional trapped run-and-tumble particles with homogeneous swimming speed and tumbling 
rate trapped in an external potential $U(x)$. Without the trapping  effects the effective temperature description results are valid. We present 
in Sec. \ref{section4} our final comments and concluding remarks.

\section{\label{section2} Theoretical considerations}
A rich variety of patterns of active motion are observed in nature, and many different mathematical 
models have been introduced to describe their dynamics. A recurrently mathematical model used to describe the motion of 
biological organisms, which takes into account the \emph{persistence} of motion ---characteristic of the active one--- corresponds to the 
random walk and its variants \cite{BergBook2,CodlingJRoySocInterface2008}. One pattern of motion that has received particular attention is called \emph{run-and-tumble} 
\cite{CodlingJRoySocInterface2008}, observed, for instance, in the motion of bacteria such as \emph{Escherichia Coli} \cite{BergBook}. The organism use cilia synchronization 
to move 
approximately in straight line and with constant speed for a random period of time (of the order of seconds), called the \emph{running period}. Immediately after a running 
period, 
cilia get desynchronized for short periods of time (tenths of seconds) at which the bacteria tumbles. Cilia get synchronized again and the particle start a running period in a 
randomly chosen direction. Run-and-tumble dynamics can be considered a paradigm for non-Brownian diffusive motion and has been used to explore some central aspects of 
nonequilibrium dynamics, such as the origins of motility-induced phase separation in systems without detailed balance \cite{CatesAnnRevCMP2015}. 

The dynamics of run-and-tumble particles consists of a particle running at a constant speed $v$, allowed 
to randomly change its direction of motion at a constant tumbling rate 
$\alpha$. The dynamics simplifies in one dimension where only two directions of motion are possible. If the speed of the particle and the 
rate of change of direction are constants, then the process is well described by the so-called telegrapher's equation, which captures in an exact manner this dynamics 
\cite{GoldsteinQJMAM1951}. A biased motion can be analyzed straightforwardly if the values of the particle speed and/or the transition rates depend on the direction of 
motion, namely $v_{R}$, $\alpha_{R}$ when the particle moves to the right and 
$v_{L}$, $\alpha_{L}$ when it moves to the left. This set of parameters embodies the description of the one-dimensional run-and-tumble dynamics. A generalization of this model 
considers the coupling of the particle's motion parameters to the environment causing gradients in the particle mobility properties  
\cite{SchnitzerPRE1993,RazinPRE2017a,RazinPRE2017b} or,
alternatively, in a mean-field description, it considers the coupling to the local population density that takes into account many-body 
effects \cite{TailleurPRL2008}. In any case such a coupling makes $\alpha$ and $v$ depend intrinsically on the particle's position $x$. Thus the 
probability densities of being 
at $x$ at the instant $t$ and moving to the right, $P_R(x,t)$, and to the left, $P_L(x,t)$, satisfy the equations
\begin{widetext}
\begin{subequations}\label{RT-Eqs}
\begin{align} 
 \frac{\partial}{\partial t}P_{R}(x,t)+\frac{\partial}{\partial x}v_{R}(x)P_{R}(x,t)&= \frac{1}{2}\left[\alpha_{L}(x)P_{L}(x,t)-\alpha_{R}(x)P_{R}(x,t)\right],\\
 \frac{\partial}{\partial t}P_{L}(x,t)-\frac{\partial}{\partial x}v_{L}(x)P_{L}(x,t)&= \frac{1}{2}\left[\alpha_{R}(x)P_{R}(x,t)-\alpha_{L}(x)P_{L}(x,t)\right],
 \end{align}
\end{subequations}
\end{widetext}
where the $v$'s and $\alpha$'s are positive functions of the particle position. Equations \eqref{RT-Eqs} can be written in an equivalent form in terms of the coarse-grained 
probability density $P(x,t)=P_{R}(x,t)+P_{L}(x,t)$ and the corresponding probability current $J(x,t)=v_{R}(x)P_{R}(x,t)-v_{L}(x)P_{L}(x,t)$, both related by the continuity 
equation 
\begin{align}\label{Continuity}
 \frac{\partial}{\partial t}P(x,t)+\frac{\partial}{\partial x}J(x,t)=0.
\end{align}
$P(x,t)$ gives the probability density of finding a particle in a position $x$ at a time $t$ independently of the direction of motion, while $J(x,t)$ gives the net flux of an ensemble of particles at $x$ and $t$, which satisfies the equation 
\begin{widetext}
\begin{equation}\label{FluxJ}
\frac{\partial}{\partial t}J(x,t)-v_{\text{rel}}(x)\frac{\partial}{\partial x}J(x,t)+\left[\alpha(x)+\gamma(x)\right]J(x,t)= 
\alpha(x)\left[\mathcal{V}_{\text{drift}}(x)P(x,t)-\frac{\partial}{\partial x}\mathcal{D}(x)P(x,t)\right].
\end{equation}
\end{widetext}
The rate of change in time of $J(x,t)$ is given on the one hand, by the advection term [second in the left-hand side of \eqref{FluxJ}] with velocity field of the probability 
flow
\begin{equation}
 v_{\text{rel}}(x)=v_{R}(x)-v_{L}(x);
\end{equation}
by the reaction term [third in the left-hand side of \eqref{FluxJ}], with reaction rate $\alpha(x)+\gamma(x)$, where
\begin{equation}
 \alpha(x)=\frac{1}{2}[\alpha_{R}(x)+\alpha_{L}(x)]
\end{equation}
is the arithmetic average of the tumbling rates, which being positive, makes $J(x,t)$ diminish in time due to the scattering process of the direction of motion, while
\begin{equation}
  \gamma(x)=\frac{v^{\prime}_{R}(x)v_{L}(x)-v^{\prime}_{L}(x)v_{R}(x)}{v_{R}(x)+v_{L}(x)}
\end{equation}
takes into account the spatial dependence of the $v$'s. 

In the right-hand side of \eqref{FluxJ}, the local drift velocity, $\mathcal{V}_{\text{drift}}(x)$, is given by 
\begin{multline}\label{DriftVel}
\mathcal{V}_{\text{drift}}(x)=\frac{\alpha_{L}(x)v_{R}(x)-\alpha_{R}(x)v_{L}(x)}{\alpha_{R}(x)+\alpha_{L}(x)}\\
+\mathcal{D}(x)\frac{d}{dx}\ln\left[\frac{\alpha_{R}(x)+\alpha_{L}(x)}{v_{R}(x)+v_{L}(x)}\right]
\end{multline}
which originates, on the one hand, in the asymmetry of the spatial dependence of the left-right transition rates and left-right moving velocities [first term in the right 
hand side of Eq. 
\eqref{DriftVel}], which vanishes when $v_{R}(x)/\alpha_{R}(x)=v_{L}(x)/\alpha_{L}(x)$. On the other hand, in a term that is proportional to the gradient of 
$\ln[(\alpha_{R}+\alpha_{L})/(v_{R}+v_{L})]$, where 
\begin{equation}\label{DiffusionRTP}
 \mathcal{D}(x)=\frac{v_{R}(x)v_{L}(x)}{\alpha(x)}.
\end{equation}
is a position-dependent diffusion coefficient \cite{TailleurPRL2008} that emerges from the random change of the particle's direction of motion, as can be deduced from the 
second term in the right-hand side of Eq. \eqref{FluxJ}, which gives the contribution to the current due to 
inhomogeneity of the probability density.

For the case of run-and-tumble particles swimming in an aqueous environment, the effects of thermal fluctuations on the particle motion might not be 
neglected. To incorporate them the total probability current  $J_{T}(x,t)$ considers the standard Fick-like probability current $-D_{T}\partial P_{T}(x,t)/\partial x$ that 
describes thermal diffusion characterized by the diffusion coefficient $D_{T}$ plus the convolution of the Gaussian propagator of thermal diffusion, 
$G_{D_{T}}(x,t)=\exp\{-x^{2}/4D_{T}t\}/\sqrt{4\pi D_{T}t} $, with the probability current that describes active motion $J(x,t)$ given in \eqref{FluxJ}. The 
total probability density of the particles positions $P_{T}(x,t)$ must be written as the convolution of the probability density contribution from active motion $P(x,t)$ given 
in \eqref{Continuity} with $G_{D_{T}}(x,t)$. We focus our analysis on the dynamics of active motion; on the one hand our analysis is exact in the regime for which the ratio 
$D_{T}/(v_{0}^{2}/\alpha_{0})\ll1$ ($v_{0}$ and $\alpha_{0}$ being the characteristic parameters of active motion); otherwise, the effects of thermal fluctuations can be 
taken into account easily as has just been explained.

The general equations \eqref{Continuity} and \eqref{FluxJ}, whose boundary conditions will be specified later, are our starting point. In order to 
connect the nonequilibrium nature of active motion to concepts used in standard nonequilibrium statisical mechanics, we consider the \emph{ideal} situation for which active 
motion is intrinsically described by constant swimming speed and constant tumbling rate. From this, we unveil a map that elucidate the connection between the nonequilibrium 
nature of active motion under trapping potentials.

\subsection{The diffusive limit: Free active motion and the emergence of effective temperature}
The simplified situation of run-and-tumble particles moving freely in a \emph{uniform source of activity}, i.e., in medium that serves as a uniform source that keeps the 
tumbling rates, 
and the right and left velocities, equal and space independent, i.e., $\alpha_R(x)=\alpha_{L}(x)=\alpha$ 
and $v_{R}(x)=v_{L}(x)=v$, can be understood analogously with the uniform 
temperature bath that causes Brownian motion. For systems in one dimension, these considerations lead to the simplest model of \emph{persistent motion} taken into account by 
the one-dimensional telegrapher's equation \cite{BourretCJP1960, BourretCJP1961, GoldsteinQJMAM1951}, which can be obtained straightforwardly from Eqs. \eqref{Continuity} and 
\eqref{FluxJ} with natural boundary conditions, namely that at $x\rightarrow\pm\infty$ the probability density and the probability current vanish. This 
generalizes the diffusion equation in that it properly accounts for the finite speed signal propagation that results in a non-Gaussian probability density function of 
the particle positions $P(x,t)$. Such a diffusion process is nonstationary and is characterized by ballistic motion in the short-time regime and normal diffusion in the 
long-time limit, where $P(x,t)$ asymptotically approximates the Gaussian solution of the diffusion equation, and a uniform effective diffusion coefficient is apparent, namely 
$D_{\text{free}}=v^2/\alpha$ \cite{GoldsteinQJMAM1951,PalacciPRL2010}. Thus, assuming uniform swimming velocities and tumbling rates leads, in the long-time limit, to a
uniform effective temperature, 
\begin{equation}\label{T0definition}
 T_{0}=\frac{v^{2}}{\alpha\, \mu\, k_{B}},
\end{equation}
if a kind of Einstein relation is assumed, where $\mu$ is a uniform parameter that describes the coupling of the particle to the fictitious 
heat bath at uniform temperature $T_{0}$.

\subsection{Trapped active motion: Stationary solutions} We are interested in the physical situations for which stationary solutions, 
$P_{\text{st}}(x)$, with vanishing probability current in a finite region exist, as happens, for instance, if the particles are trapped either by impenetrable 
walls or by some external forces. It is well known that the persistence effects of run-and-tumble particles makes the particles to explore the container walls if the 
persistence length is larger than or comparable  to the confinement length \cite{FilySoftMatt2014}. This is the case also if 
confinement is due to energetic constraints, for instance, when particles are trapped by an external potential. In this case, stationary solutions are obtained from Eqs. 
\eqref{Continuity} and \eqref{FluxJ} by setting  $J(x,t)=0$ for all positions in the spatial region allowed by the confinement, and
this leads to
\begin{equation}\label{eq_current}
 \mathcal{V}_{\text{drift}}(x)P_{\text{st}}(x)-\frac{d}{dx}\mathcal{D}(x) P_{\text{st}}(x)=0.
\end{equation}

Notice that position dependence of the $v$'s and of the $\alpha$'s is a necessary but not sufficient condition for the existence of stationary solutions. Further, the 
relations \eqref{DriftVel} and \eqref{DiffusionRTP} lead to a mapping between the stationary solutions of run-and-tumble particles and the corresponding 
 stationary distributions of a drift-diffusion equation of Brownian motion in inhomogeneous media, analogously to the one presented in Ref. 
\cite{TailleurPRL2008}. This mapping will be exploited to give a precise interpretation of confined active motion as a nonequilibrium feature as will be unveiled in the 
following sections.

The solutions to Eq. \eqref{eq_current} describe the flux-free steady states that can be written, by considering the nonlocality of the ratio 
$\mathcal{V}_{\text{drift}}(x)/\mathcal{D}(x)$ as \cite{SchnitzerPRE1993,TailleurPRL2008,BechingerRMP2016}
\begin{equation}\label{GralSolution}
 P_{\text{st}}(x)=\frac{N}{\mathcal{D}(x)}\, \exp\left\{\int^{x}dx^{\prime}\frac{\mathcal{V}_{\text{drift}}(x^{\prime})}{\mathcal{D}(x^{\prime})} \right\},
\end{equation}
where $N$ is the required normalization factor \footnote{It can be shown that the stationary solution is normalizable under generic physical 
situations. For the purpose of our analysis, it is enough to requiere  $v_{R}(x)$ to be a bounded monotonous decreasing function of the particle position, vanishing 
at $x_{\text{max}}$ which defines the right boundary. Analogously,  $v_{L}(x)$ must  be a bounded monotonous increasing function that start rising from the value zero at 
$x_{\text{min}}$ that defines the left boundary.} that has units of length over time and the symbol $\int^{x}dx^{\prime}\, f(x^{\prime})$ denotes the antiderivative 
or indefinite integral of $f(x)$. Clearly, such solutions do not comply with the well-known Boltzmann-Gibbs factor that describes thermodynamic equilibrium 
(homogeneous temperature), and therefore the set of solutions \eqref{GralSolution} can be interpreted as a 
novel class of nonequilibrium stationary distributions known as \emph{superstatics} proposed in Ref. \cite{BeckPRE2005}. Although the original concept of superstatics 
encloses the 
non-Boltzman-Gibbs stationary distributions that emerge from the superposition of local Boltzmann-Gibbs factors \cite{BeckPRE2005,ChavanisPhysicaA2006}, the non-Boltzmannian 
stationary distribution given by Eq. \eqref{GralSolution} emerges as consequence of a spatial dependence of the kinematic parameters, namely the $v$'s and  the $\alpha$'s. 
As will be shown in the following lines, expression \eqref{GralSolution} can be mapped to the class of stationary non-Boltzmannian distributions of Brownian 
motion in a given random inhomogeneous medium \cite{ButtikerZPhysB1987,SanchoPRE2015}. By pointing out the distinction between these two wide classes of non-Boltzmannian 
stationary distributions, we attempt to elucidate the organization of a rather small part of the vast nonequilibrium stationary distributions that occur in an enormous number 
of domains of physics. 

Notice, however, that if the parameters of active motion satisfy certain conditions, that is, if the coupling between the particle motion and the environment are devised 
such 
that $\mathcal{D}(x)=D$ is independent of the particle's position and $\mathcal{V}_\text{drift}(x)=-G^{\prime}(x)$ caused by the 
pseudo potential $G(x)$ is an arbitrary function of $x$ (this can always be done in one dimension), 
then $P_{\text{st}}(x)$ given in \eqref{GralSolution} can be written as the Boltzmann-Gibbs-like weight,
\begin{equation}
 P_{\text{B-G}}(x)=Z^{-1}(D)\exp\left\{-G(x)/D\right\},
\end{equation}
where $Z(D)=\int_{-\infty}^{\infty}dx^{\prime}\,\exp\left\{- G(x^{\prime})/D\right\}$ is reminiscent of the single-particle partition function of the canonical ensemble and 
$D$ a homogeneous global parameter that can be related to an effective thermodynamic quantity, like the effective temperature $T_{0}$ as has been demonstrated experimentally 
and theoretically \cite{CugliandoloPRE1997,TailleurPRL2008,TailleurEPL2009,PalacciPRL2010,FodorPRL2016}. The corresponding free energy, according to the theory 
of stochastic thermodynamics \cite{SeifertRepProgPhys2012}, is given by $F=-D\ln Z(D)$. If this is the case, then the fluctuation-dissipation theorem is expected to be valid 
and 
the time-reversal symmetry of the microscopic dynamics is guaranteed, and thus the ratio of the transition rate from position $x$ to $x^{\prime}$, 
$W(x,x^{\prime})$, to the inverse process $W(x^{\prime},x)$, is 
given by the well-known equilibrium result $\exp \left\{-[G(x^{\prime})-G(x)]/D\right\}$.

Therefore, we can conclude that the intrinsic nature of the non-Boltzmann-Gibbs equilibrium solutions of active motion resides in the fact that no such homogeneous, global 
parameter can be identified, thus leading to a situation that is intrinsically out of thermodynamic equilibrium. This statement is made clear by recognizing that the solution 
given in \eqref{GralSolution} can be formally interpreted as the stationary solution of an overdamped, passive Brownian 
particle that diffuses with constant mobility $\mu$ in a fictitious medium of inhomogeneous temperature $\mathcal{T}(x)$ 
\cite{ButtikerZPhysB1987,MatsuoPhysicaA2000,DurangPRE2015,SanchoPRE2015} under the influence of an effective external potential 
$\mathcal{U}(x)$ if the following identifications are made:
\begin{subequations}\label{Identifications}
 \begin{align}
      \label{TempProfile}
      \mathcal{T}(x)&=\mathcal{D}(x)/\mu\, k_{B},\\
      \mathcal{U}^{\prime}(x) &=-\, \mathcal{V}_{\text{drift}}(x)/\mu.
 \end{align}
\end{subequations}

In the context of Brownian passive particles, it is well known that inhomogeneous temperature profiles have a profound consequence on the local stability of the 
stationary solutions, known as the Landauer effect \cite{LandauerPRA1975, vanKampenIBMJRD1988}. Certainly, a hot layer can change the relative stability of 
equilibrium points of a particle diffusing in a bistable potential. This observation might have important applications, namely, by properly 
choosing a spatial inhomogeneity of temperature [namely Eq. \eqref{TempProfile}], it is possible to obtain a desired 
stationary distribution of particles, as, for example, to mimic the persistence effects of active motion of biological organisms by the diffusion of  trapped passive 
particles.

Note,  first, that the physical assumptions underlying the establishment of relations \eqref{Identifications} indicates that two elements of
nonequilibrium can be identified; the immediate one refers to the inhomogeneity of the effective temperature, and the other, which is more subtle, refers to the uniformity of 
the 
mobility $\mu$. Indeed, this last element contrasts with the case when \emph{only} conditions of \emph{local equilibrium} are assumed, for which the mobility of an 
overdamped Brownian particle is space dependent and constrained to the spatial dependence of temperature in order to satisfy a local fluctuation-dissipation relation 
\cite{vanKampenIBMJRD1988,WidderPhysicaA1989}. Though precise information regarding the dissipative coupling of the particle's motion to the medium is required to avoid any 
loose assumption, in the present analysis a space-independent mobility is assumed. 

\section{\label{section3}Application to a Specific case: trapped run-and-tumble particles in an external potential}

Although the coupling of the particle's motion to the medium is in general complex, as when chemotaxis behavior 
is considered, here we analyze the simple situation for which a particle swims at constant speed $v$ in a viscous fluid at low Reynolds numbers, such that an 
overdamped dynamics is valid. We also assume that the particle tumbles symmetrically at constant rate $\alpha$ and that the motion is restricted by an external trapping 
potential $U(x)$. Under these considerations the effective right and left swimming speeds become space dependent and are given by 
\begin{subequations}\label{SwimmingSpeeds}
\begin{align}
v_{R}(x)&=v-\mu\, U^{\prime}(x),\label{RightSwimmingSpeed}\\
v_{L}(x)&=v+\mu\, U^{\prime}(x),\label{LeftSwimmingSpeed}
\end{align}
\end{subequations}
where $\mu$ is the mobility of the particle in the fluid, which for simplicity is assumed to be space independent and finite. With these considerations we have that Eqs. \eqref{Identifications} can be rewritten as 
\begin{subequations}\label{VdriftEffectiveTemp}
 \begin{align}
      \mathcal{V}_{\text{drift}}(x)&=-\mu\, U^{\prime}(x),\\
      \mathcal{T}(x)&=T_{0}\left\{1-\frac{\mu^{2}}{v^{2}}\left[U^{\prime}(x)\right]^{2}\right\},\label{TempProfileParticular}
 \end{align}
\end{subequations}
where the effective temperature $T_{0}$ has been introduced in Eq. \eqref{T0definition}. It is clear that in the \emph{diffusive limit}, $v\rightarrow\infty$, 
$\alpha\rightarrow\infty$ such that $v^{2}/\alpha=\mu k_{B}T_{0}$ is kept constant \cite{TailleurEPL2009}, we recover the uniform effective temperature 
description given by Eq. \eqref{T0definition}, since in such a limit the ratio of the drift velocity (caused by the external potential) to 
the particle swimming velocity vanishes. Also in this 
limit, the persistence length, $l_{\text{pers}}\equiv v/\alpha$, that characterizes the length scale of fluctuations, goes to zero, thus satisfying a kind of 
fluctuation-dissipation relation that guaranties equilibrium states characterized by a uniform, effective temperature $T_{0}$ \cite{DieterichNaturePhys2015}.

In the persistent regime, on the other hand, the motion of the particle is sharply bounded by the external potential. Indeed, the particle cannot swim beyond a 
characteristic distance $x_{\text{max}}$ from the local stable point of the trapping potential, where the self-propulsive 
force $\mu^{-1}v$ equals that of the trapping force $-U^{\prime}(x)$, i.e., $x_{\text{max}}$ is defined by the position at which the right speed $v_{R}$ 
vanishes, the solution of the equation $\mu\left\vert U^{\prime}(x_{\text{max}})\right\vert= v$. Thus, in the case of an even-symmetric potential $[U(-x)=U(x)]$ with a unique 
global minimum at the origin, the particle moves in the region of space defined by $[-x_{\text{max}},x_{\text{max}}]$, where 
$-x_{\text{max}}$ is the position at which $v_{L}$ vanishes. Notice now that the inhomogeneous temperature \eqref{TempProfileParticular} takes its maximum value, $T_{0}$, 
just 
at the positions corresponding to the minima of the trapping potential $U(x)$ and vanishes at the positions $\pm x_{\text{max}}$. The fictitious medium, 
being ``hotter'' at the potential minima, ``push out'' the particles from the corresponding stable positions of $U(x)$ toward the new stable positions given by the local 
minima of $U_{\text{eff}}(x)$ on the interval $[-x_{\text{max}},x_{\text{max}}]$, thus changing the system stable points of the trapping potential for those at the boundaries 
(a similar effect was analyzed by Landauer for the case of a simple hotter layer in between the local minima of a bistable potential 
\cite{LandauerPRA1975} and extended by van Kampen for a general potential and a general temperature inhomogeneity \cite{vanKampenIBMJRD1988}). This gives a precise picture 
that pinpoints the nonequilibrium nature of the stationary distribution of trapped run-and-tumble particles.

The probability distribution \eqref{GralSolution} can be written as
\begin{equation}\label{EquilSol}
P_{\text{st}}\left(x\right)=\mathcal{Z}^{-1}\exp\left\{-\int^{x} dx'\, \frac{U_{\text{eff}}^{\prime}(x^{\prime})}{k_{B}\mathcal{T}\left(x'\right)}\right\},
\end{equation}
where the effective potential
\begin{equation}\label{EffectivePotential}
 U_{\text{eff}}(x)=U(x)+k_{B}\mathcal{T}(x)
\end{equation}
takes into consideration the appearance of the fictitious thermophoretic force $-k_{B}\mathcal{T}^{\prime}(x)$ \cite{WidderPhysicaA1989}, due to the spatial inhomogeneity of 
the effective temperature and explicitly given by $-k_{B}\mathcal{T}^{\prime}(x)=(2k_{B}T_{0}\mu^{2}/v^{2})U^{\prime}(x)U^{\prime\prime}(x)$. 

The stationary distributions given by \eqref{EquilSol} form a particular class of the 
\emph{superstatistics} given by \eqref{GralSolution} \cite{ButtikerZPhysB1987,SanchoPRE2015}, whose nonlocal character allows us to interpret it by writing the 
indefinite integral in the argument of the exponential as a Riemann sum starting at $x_{\text{min}}$, for instance, as a product of locally Boltzmann-Gibbs factors. In the 
diffusive limit the nonlocal character of \eqref{EquilSol} vanishes and the Boltzmann-Gibbs statistics is recovered as a particular case. The normalizing 
constant, $\mathcal{Z}$, corresponds to the system partition function, which is given explicitly by 
\begin{equation}\label{PartitionFunction}
\mathcal{Z}= \int dx\, \exp\left\{-\int^{x} dx'\, \frac{U_{\text{eff}}^{\prime}(x^{\prime})}{k_{B}\mathcal{T}\left(x'\right)}\right\}.
\end{equation}
Though no homogeneous effective temperature exists, a local free energy density $\mathcal{F}(x)$ can be defined through the relation 
\begin{equation}
 \mathcal{Z}=\exp\left\{-\int dx^{\prime}\frac{\mathcal{F}(x^{\prime})}{k_{B}\mathcal{T}(x^{\prime})}\right\},
\end{equation}
which takes into account the nonlocal effects due to the inhomogeneity of the medium. The detailed discussion of this aspect will be presented elsewhere. 

The effects of the thermophoretic potential, $k_{B}\mathcal{T}(x)$, that incorporates the persistence of active motion in an effective manner are conspicuously revealed in 
the  stability nature of the equilibrium positions of $U_{\text{eff}}(x)$. In the diffusive limit, the effective temperature \eqref{TempProfileParticular} becomes spatially 
uniform and therefore the equilibrium positions of $U_{\text{eff}}(x)$ corresponds to those of $U(x)$. In this situation, the particles accumulate around the stable positions 
(global minima) of the trapping potential $U(x)$. Furthermore, in the same limit, the Boltzmann-Gibbs distribution $P_{B-G}(x)$, which describes the thermodynamic 
equilibrium, 
emerges from  expression \eqref{EquilSol} \cite{Solon2015EPJST2015}. As the effects of persistence become more apparent, new equilibrium positions besides those of $U(x)$ 
appear in the system. These new equilibrium positions are explicitly given as the solutions of the equation $U^{\prime\prime}(x)=\alpha/(2\mu)$, which explicitly exhibits the 
dependence on the ratio of the inverse of the persistence time and the mobility. The explicit appearance of the second derivative requires an external potential that varies 
rapidly enough with position in order to have new equilibrium positions other than those given by the minima of $U(x)$. 

On the other hand, the sharp accumulation of particles at the confining boundaries, which is nonequilibrium hallmark of dilute active systems, appears just in the persistent 
regime, when the persistence length $l_{\text{pers}}$ is larger or of the order of the characteristic length of confinement. The departure from the well-known 
equilibrium Boltzmann-Gibbs distribution can be identified by rewriting Eq. \eqref{EquilSol} as 
\begin{multline}\label{EquilSol2}
 P_{\text{st}}\left(x\right)=\mathcal{Z}^{-1}\exp\left\{-\frac{U_{\text{eff}}(x)}{k_{B}T_{0}}\right\}\times\\
 \exp\left\{-\sum_{l=1}^{\infty}\left(\frac{\mu}{v}\right)^{2l}\int^{x}dx^{\prime}\frac{\left[U^{\prime}(x^{\prime})\right]^{2l+1}}{k_{B}T_{0}}\right\},
\end{multline}
where the Boltzmann-Gibbs factor, with the effective potential $U_{\text{eff}}(x)$, is explicitly factorized and expression \eqref{TempProfileParticular} has been explicitly 
used. Clearly, the stationary Boltzmann-Gibbs distribution is recovered in the diffusive limit. In the other extreme limit, when the thermophoretic potential dominates, the 
stationary distribution that describes the sharp accumulation at the boundaries goes asymptotically as
\begin{equation}
 P_{\text{st}}(x)\propto\frac{v}{k_{B}T_{0}\mu}\left\lbrace1-\frac{\mu^{2}}{v^{2}}\left[U^{\prime}(x)\right]^{2}\right\rbrace^{-1}.
\end{equation}

Although the stationary distribution given by \eqref{EquilSol} does not correspond to the one of Boltzmann and Gibbs, we maintain the use of the Boltzmann-Gibbs entropy 
$S=-k_{B}\int dx\, P_{st}(x)\, \ln P_{st}(x)$ \footnote{We have chosen this more standard approach instead of the use of generalized 
entropies --as the Tsallis entropy \cite{TsallisJStatPhys1988} or the generalized entropies of superstatistics \cite{HanelPNAS2011}-- from which such distributions are 
obtained by maximization under appropriate constraints, in an attempt to unveil the physical meaning through the stochastic thermodynamics of the non-Boltzmann-Gibbs 
distributions obtained for run-and-tumble particles.}, and with the expression for the flux-free stationary distribution in the form of \eqref{EquilSol2}, we can give an 
interpretation of the process considered under the point of view of stochastic thermodynamics. The free energy $F=-k_{B}T_{0}\ln\mathcal{Z}$ can be written 
as
\begin{equation}
 F=E_{\text{eff}}-T_{0}S,
\end{equation}
where the effective internal energy, $E_{\text{eff}}$, is defined by
\begin{multline}
  E_{\text{eff}}=\langle U(x)\rangle+\frac{1}{2}m_{\text{eff}}\, v^{2}\left[1-\frac{\left\langle\mathcal{V}_{\text{drift}}^{2}\right\rangle}{v^{2}}\right]\\
  +U^{\prime}(x_{\text{max}})\sum_{l=1}^{\infty}\left(\frac{v_{m}}{v}\right)^{2l}\left\langle\int^{x}dx^{\prime}\,\left[\frac{U^{\prime}(x^{\prime})}{U^{\prime}(x_{\text{max}})}\right]^{2l+1}\right\rangle,
\end{multline}
with $m_{\text{eff}}=2/\mu\alpha$ and $\langle\cdot\rangle$ denotes the average using the stationary distribution $P_{\text{st}}(x)$. Notice that the first two terms in the 
right-hand side of the last equation contribute to standard mechanical energy since the second term, being positive definite, can be interpreted as an effective kinetic 
energy. The last term gives the contribution from the effects of persistence in series expansion in powers of the dimensionless parameter $v_{m}/v$, $v_{m}$ given by $\mu\, 
U^{\prime}(x_{\text{max}})$. In the diffusive limit we have simply that $E_{\text{eff}}=\langle U(x)\rangle_{\text{B-G}}$, where $\langle\cdot\rangle_{\text{B-G}}$ denotes 
the average taken with the Boltzmann-Gibbs distribution.

\paragraph*{Numerical simulations.-}
As stated in the previous section, the stationary behavior of trapped active motion, contained in expression \eqref{GralSolution} and particularly in \eqref{EquilSol}, can be
regarded as the stationary behavior of a passive Brownian particle that moves in an inhomogeneous temperature medium. Thus, for the cases of interest, stationary realizations 
of the particle trajectories can be obtained directly from the Langevin equation \cite{MatsuoPhysicaA2000,DurangPRE2015,SanchoPRE2015}  
\begin{equation}\label{eq_langevin}
 \frac{d}{dt}x(t)=-\mu  U^{\prime}(x)+\sqrt{\mathcal{T}(x)}\, \xi(t),
\end{equation}
if initial conditions are compatible with the stationary solution \eqref{EquilSol}. In Eq. \eqref{eq_langevin}, $\xi(t)$ denotes Gaussian white noise, with vanishing mean
$\langle\xi(t)\rangle=0$ and autocorrelation function $\langle\xi(t)\xi(t')\rangle=2\mu k_{B}\delta(t-t')$ and $\mathcal{T}(x)$, the inhomogeneous medium 
temperature \eqref{TempProfileParticular} that encodes the features of active motion. Ensemble calculations of realizations of these trajectories lead 
to all the observable quantities of interest, in particular those quantities of interest in \emph{stochastic thermodynamics} (see, for instance, the review 
\cite{SeifertRepProgPhys2012}).

We apply the ideas developed up to now to particular realizations of the potential $U(x)$ that had been considered before in the literature, both theoretically and 
experimentally, namely the linear potential in the sedimentation process of active particles \cite{TailleurPRL2008,TailleurEPL2009,PalacciPRL2010,EnculescuPRL2011}, the 
harmonic potential trapping active particles \cite{TailleurEPL2009,TakatoriNatcomms2016}, and the diffusion of active particles in a double well-potential 
\cite{CapriniJChemPhys2019}.

\subsection{\label{Sedimentation}Sedimentation: Linear potential $U(x)=mgx$ }
The simplest physical realization for run-and-tumble particles in an external potential, corresponds to the case when $U(x)$ is linear with distance, i.e., when the particles 
are subject to a constant force, as is the case of active particles that swim above a hard wall in the presence of the gravitational force $-mg$. The probability density of 
finding a particle at height $x$ above the wall has been found in one dimension for run-and-tumble particles \cite{SchnitzerPRE1993, TailleurPRL2008} and in higher dimensions 
in Ref.~\cite{TailleurEPL2009}; further, this situation has been realized experimentally in three dimensions for active Brownian particles \cite{PalacciPRL2010}. 

From relations \eqref{VdriftEffectiveTemp} we have that the drift velocity $\mathcal{V}_{\text{drift}}=-\mu mg$ corresponds to the sedimentation velocity $-v_{\text{sed}}$
, and the effective temperature $\mathcal{T}(x)=T_{0}(1-v_{\text{sed}}^{2}/v^{2})$ is spatially homogeneous.  The familiar exponential decrease of the 
probability density 
with the distance from the wall is recovered from Eq. \eqref{EquilSol},
\begin{equation}\label{EquilSolSedimentation}
 P_{\text{st}}(x)=\frac{mg}{k_{B}T_{0}(1-v_{\text{sed}}^{2}/v^{2})}e^{- mgx/k_{B}T_{0}(1-v_{\text{sed}}^{2}/v^{2})},
\end{equation}
The uniformity of the effective temperature allows us to interpret Eq. \eqref{EquilSolSedimentation} as the probability density of a Brownian particle diffusing under the 
effects of the gravitational force in a medium of homogeneous temperature $T_{0}(1-v_{\text{sed}}^{2}/v^{2})$ \cite{TailleurPRL2008,TailleurEPL2009,PalacciPRL2010}, with 
$v_{\text{sed}}< v$. By self-propelling, active particles develop higher speeds to overcome $v_{\text{sed}}$. If $v_{\text{sed}}/v\ll1$  the effective temperature 
corresponds to the maximum value $T_{0}$. In contrast, the temperature gets arbitrarily close to zero as $v$ gets arbitrarily close to $v_{\text{sed}}$, if this is the 
case, the particles 
accumulate all over the wall. 

\subsection{\label{SubSecHarmonicPotential}Power law trapping potentials: The harmonic external potential}
Experiments that analyze the effects of trapping potentials on the diffusion of active particles have been realized. In some approximation the trapping potential can be 
approximated by a harmonic potential for which analytical calculations can be devised; however, the effects of steeper potential \cite{DhararXiv1811.03808} are also of 
interest since they might approximate better the experimental trapping potential.

The analysis of the diffusion of active particles confined by a harmonic potential has been considered theoretically \cite{TailleurEPL2009,StarkEPJST2016} and, more recently, 
experimentally in a two-dimensional systems of active Brownian particles confined by transverse acoustic forces of a single-beam transducer \cite{TakatoriNatcomms2016} and in 
a two-dimensional system of an optically trapped passive Brownian particle coupled to a bath of active particles \cite{ArgunPRE0216}. 

\begin{figure}[t]
\includegraphics[width=\columnwidth]{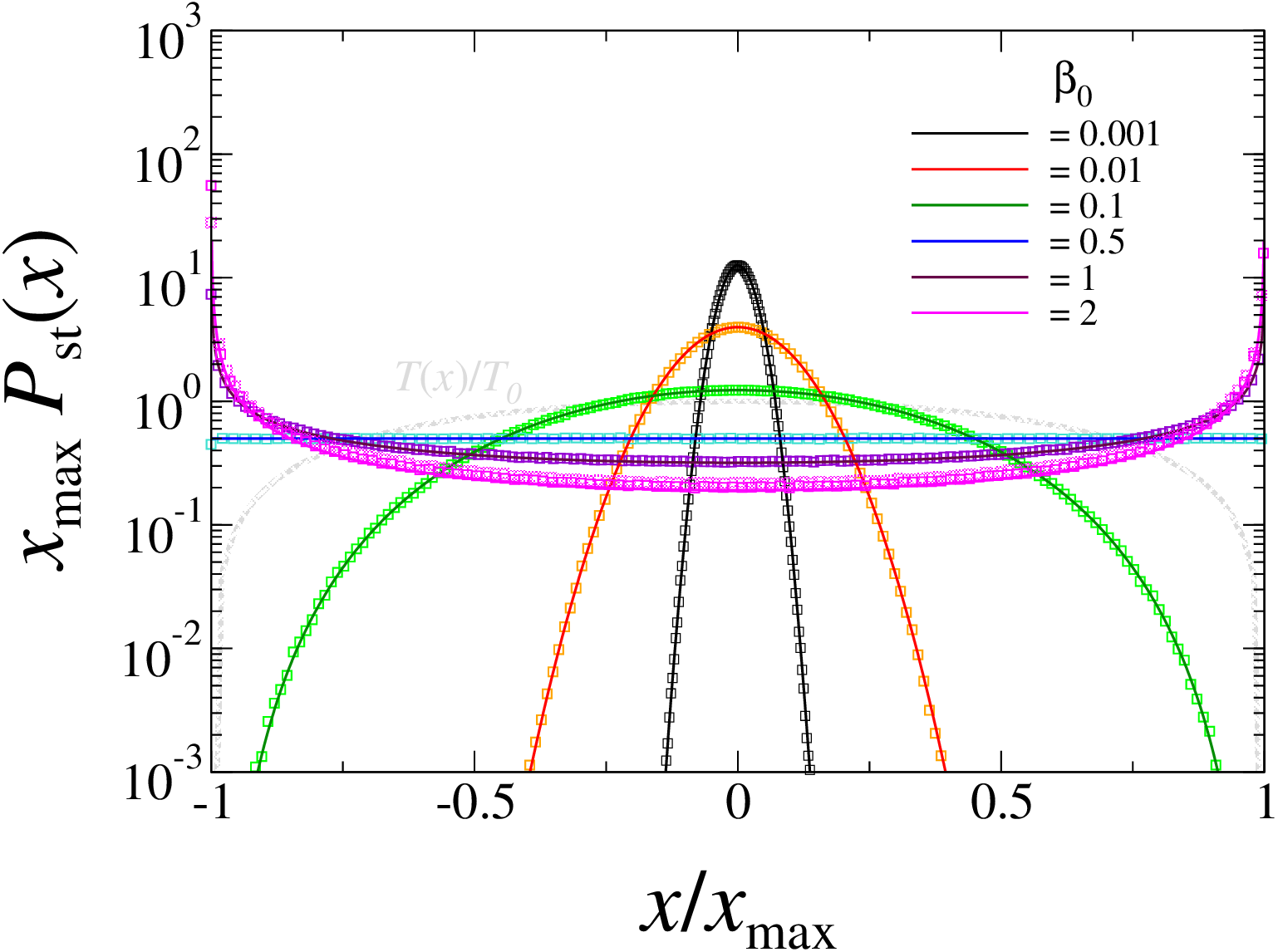}
\caption{(Color online) \label{Fig:PDF_HO} The stationary probability density $P_{\text{st}}(x)$ given by \eqref{SolStationaryHO} (solid lines) for $\beta_{0}=$ 2, 1, 0.5, 
0.1, 0.001, and 0.001 is compared with the corresponding one obtained from the numerical integration of the Langevin equation \eqref{eq_langevin} with $\mathcal{T}(x)$ given 
by 
\eqref{DB2HO}.}
\label{pdfHP}
\end{figure}

In this section we consider run-and-tumble particles trapped in the following one-dimensional harmonic trapping potential:
\begin{equation}\label{HarmonicPot}
U\left(x\right)=\frac{1}{2}\kappa_{1}\, x^{2},
\end{equation}
where $\kappa_{1}$ is a constant that characterizes the intensity of the trapping potential. For this potential the local effective temperature 
\eqref{TempProfileParticular} is
\begin{equation}\label{DB2HO}
 \mathcal{T}(x)=T_{0}\left[1-x^{2}/x_{\text{max}}^{2}\right],
\end{equation}
where $x_{\text{max}}=v/\mu \kappa_{1}$ is a length scale related to confinement, which in this case coincides with the maximum displacement for an active particle that moves 
with speed $v$. At the center of the trap, the local temperature acquires its maximum value, given by the effective temperature of free diffusion $T_{0}$ and 
vanishes at $x=\pm x_{\text{max}}$. In this case the thermophoretic force is linear in the displacement and repulsive from the center of the trap and can be written as 
$(2\kappa_{1}^{2}\mu/\alpha)\, x$. This force opposes to the one due to the harmonic potential, giving rise to the effective force $-\kappa_{1}(1-2\mu \kappa_{1} /\alpha)x$. 
The dimensionless parameter 
\begin{equation}\label{beta0HO}
 \beta_{0}=\mu \kappa_{1}/\alpha,
\end{equation}
which is equivalent to the ratio $l_{\text{pers}}/x_{\text{max}}$, measures the competition between the effects of confinement and persistence and drives the 
system into qualitatively different equilibrium distributions. Notice that $\beta_{0}$ is equivalent to the ratio of the energy $\kappa_{1}l_{\text{pers}}^{2}$ to the 
effective thermal energy $k_{B}T_{0}$. Thus, the effective potential can be written as $U_{\text{eff}}(x)=\frac{1}{2}\kappa_{1}x^{2}(1-2\beta_{0})+k_{B}T_{0},$ and from it, 
we deduce the following 
bifurcation scheme: For $2\beta_{0}<1$, $P_{\text{st}}(x)$ is unimodal around the center of the trap, which corresponds to the unique stable equilibrium position of 
$U_{\text{eff}}(x)$; for $2\beta_{0}=1$ the effects of active motion cancel out those of the trapping potential, and thus no net force acts on the particle giving rise to a 
uniform distribution into the whole interval $[-x_{\text{max}},x_{\text{max}}]$ (see Fig. \ref{pdfHP}), while for $2\beta_{0}>1$ the equilibrium position becomes unstable and 
the distribution $P_{\text{st}}(x)$ exhibits peaks at the borders $\pm x_{\text{max}}$ due to the net force from the center that pushes outward. This bifurcation is directly 
shown in 
the corresponding equilibrium solution \eqref{EquilSol2}
\begin{equation}\label{SolStationaryHO}
P_{\text{st}}(x)=\mathcal{Z}^{-1}\left[1-x^{2}\frac{\beta_{0}^{2}}{l_{\text{pers}}^{2}}\right]^{(1-2\beta_{0})/2\beta_{0}},
\end{equation}	
where the partition function $\mathcal{Z}$ is given by
\begin{equation}\label{ZHO}
\mathcal{Z}=\frac{\sqrt{\pi}\,\Gamma\left[\frac{1}{2\beta_{0}}\right]}{\Gamma\left[\frac{1}{2}+\frac{1}{2\beta_{0}}\right]}\frac{l_{\text{pers}}}{\beta_{0}}
\end{equation}
which depends explicitly on $\beta_{0}$, and $\Gamma(z)$ denotes the gamma function. It can be shown straightforwardly that in the diffusive limit, which can be stated 
equivalently 
as $\beta_{0}\rightarrow0$, $l_{\text{pers}}\rightarrow0$ such that $\beta_{0}/l_{\text{pers}}^{2}$ is finite, the Boltzmann-Gibbs equilibrium distribution 
$P_{B-G}(x)$ is recovered, i.e.,
\begin{equation}\label{BoltzmannHO}
P_{\text{B-G}}(x)=\sqrt{\frac{\beta_{0}}{2\pi l_{\text{pers}}^{2}}}\, \exp\left\{-\beta_{0}\frac{x^{2}}{2l_{\text{pers}}^{2}}\right\}.
\end{equation}

The stationary distribution \eqref{SolStationaryHO} corresponds to the class of probability distributions in statistical physics for which the Boltzmann-Gibbs distribution  
is recovered as a particular limit, in this case, in the limit $\beta_{0}\rightarrow0$. Indeed, the distribution given in Eq. \eqref{SolStationaryHO} is of the kind 
of the so-called $q$-Gaussian distribution \cite{NaudtsJPhys2010}, which is of the form 
\begin{equation}\label{QGaussian}
 \exp_{q}\left(-x^{2}\right)=\left[1-\left(\frac{1-q}{2-q}\right)^{2}x^{2}\right]^{\frac{1}{1-q}}
\end{equation}
from which the usual Gaussian function is recovered as $q\rightarrow1$. Notice that the definition of the $q$-exponential used in Tsallis statistics differs slightly from the
one given in \eqref{QGaussian}, but in terms of this one we have 
\begin{equation}
 P_{st}(x)=\mathcal{Z}^{-1}\exp_{q}\left[-x^{2}/(2l_{\text{pers}})^{2}\right],
\end{equation}
where the $q$ parameter is directly related to $\beta_{0}$ through 
\begin{equation}
 q=\frac{4\beta_{0}-1}{2\beta_{0}-1}.
\end{equation}
This last result is one of the main points of the paper, which belongs to the few systems for which the $q$ parameter is comprehensibly computed from the  
time- and length scales of the system rather than from fitting procedures.

\begin{figure}
\includegraphics[width=\columnwidth]{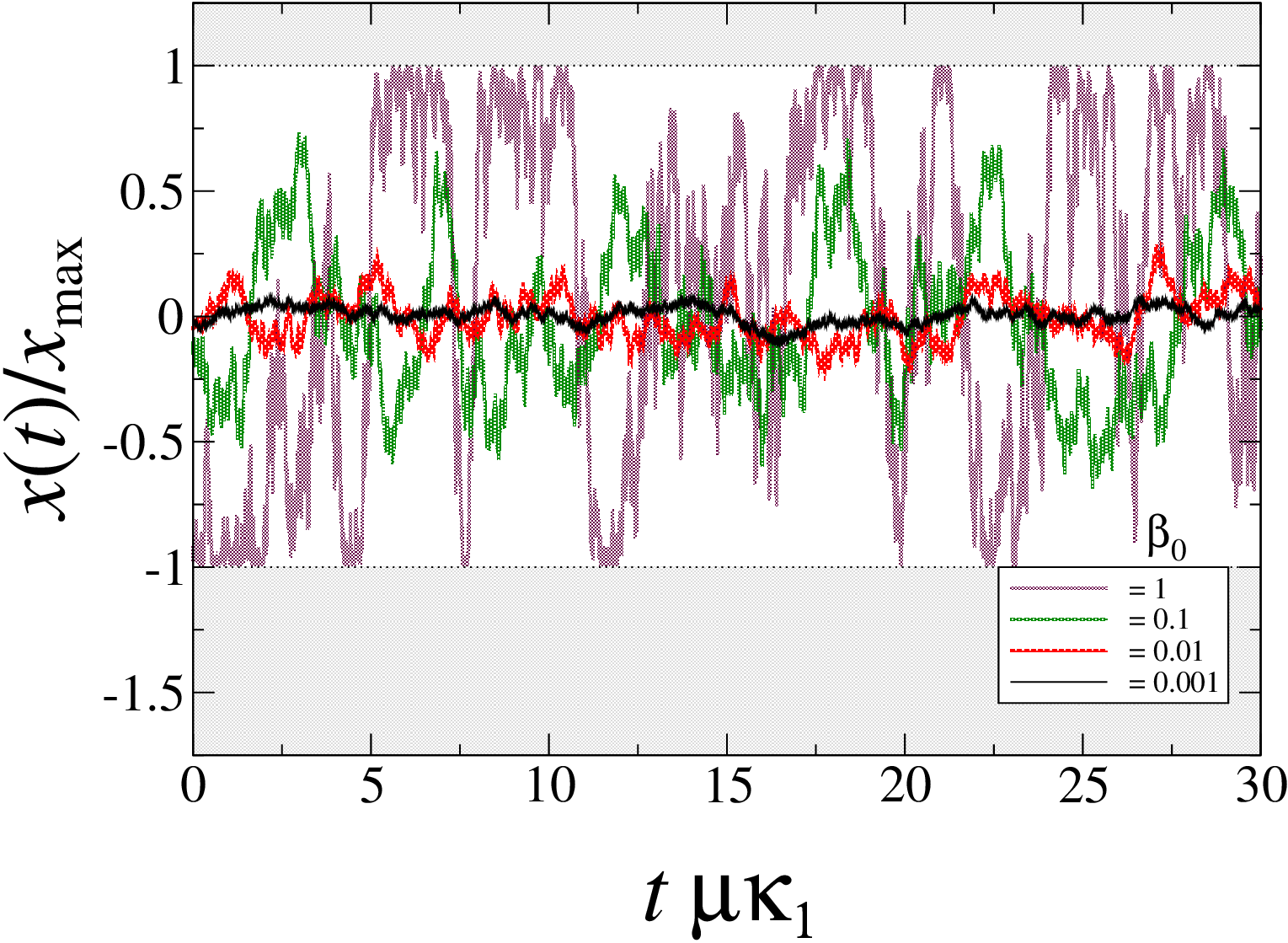}
\caption{(Color online) Trajectories in the stationary state for Brownian particles diffusing within the harmonic potential with spatially diffusion coefficient given in 
\eqref{DB2HO}, for different values of $\beta_{0}$, namely 0.001, 0.01, 0.1, and 1. Clearly, for large values of $\beta_{0}$, persistence is conspicuous and the 
particles explore more the region around the returning points $\pm x_{\text{max}}$. The shaded area marks the region of space inaccessible to the particles.}
\label{TrajectoriesHO}
\end{figure}

On the other hand, we prove the validity of Langevin dynamics given by the prescription \eqref{eq_langevin} by computing the stationary 
probability distribution \eqref{SolStationaryHO} from the ensemble average $\langle\delta[x-x(t)]\rangle_{\text{st}}$ over a set of stationary trajectories of passive 
Brownian particles diffusing in an inhomogeneous thermal bath with temperature profile $\mathcal{T}(x)$ given by Eq. \eqref{DB2HO}. In Fig. \ref{TrajectoriesHO} we present 
some of these trajectories for different realizations of $x(t)$ for a given value of the parameter $\beta_{0}$.
In Fig. \ref{pdfHP} we compare the analytical stationary distribution \eqref{SolStationaryHO} (solid lines) for different values of $\beta_{0}$, with the ones obtained from 
the numerical solutions of the Langevin equation \eqref{eq_langevin} (symbols).

\begin{figure}
 \includegraphics[width=\columnwidth]{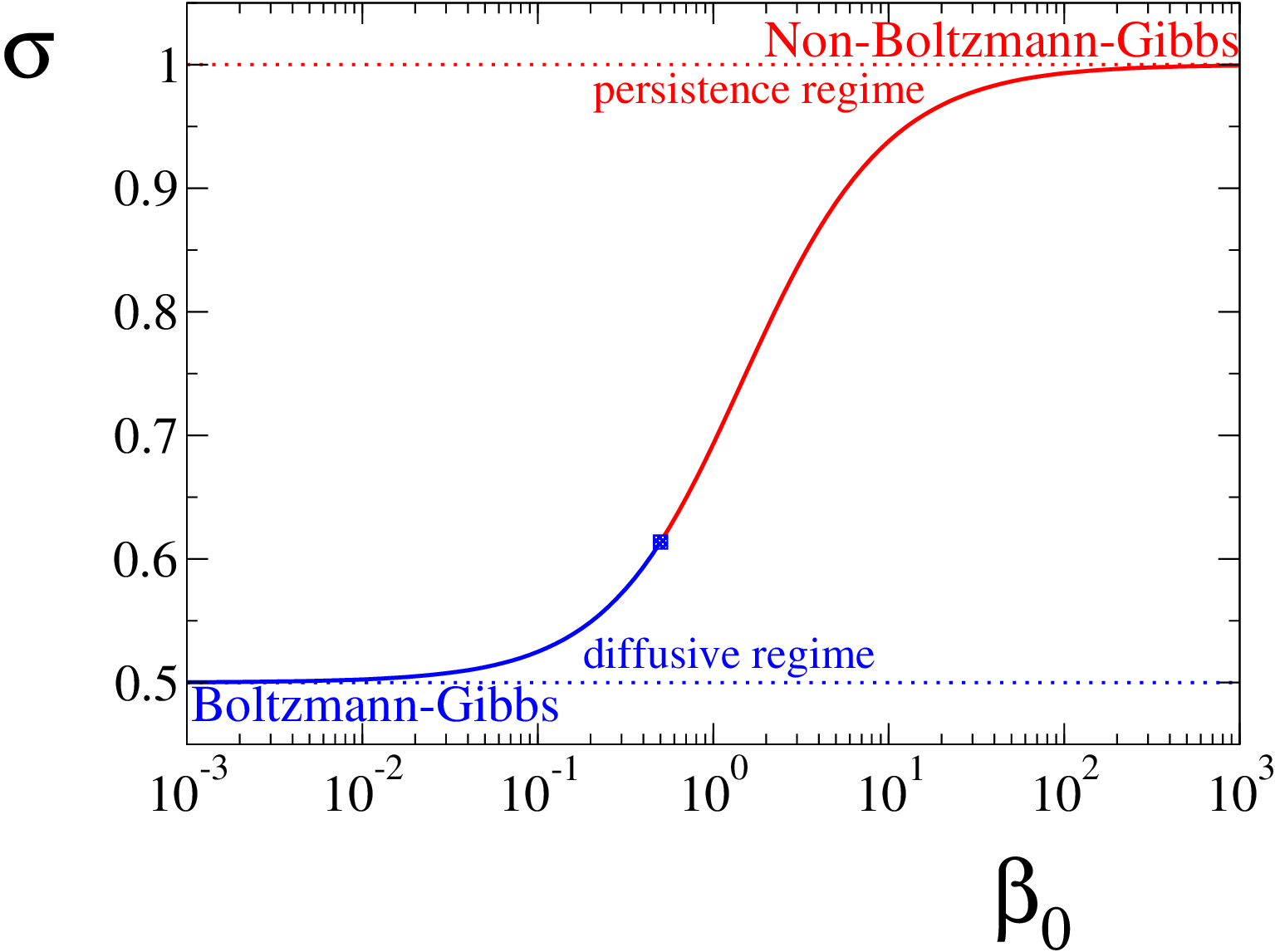}
 \caption{(Color online) $\sigma$ as function of $\beta_{0}$. It saturates in the persistent regime ($\beta_{0}\gg 1$) to the value 1, while in the 
diffusive regime ($\beta_{0}\ll1$), it characterizes the Boltzmann-Gibbs distribution with the value $\frac{1}{2}.$ The square marks the value of $\sigma$ at $\beta_{0}=1/2$ 
at which the bifurcation occurs. The probability density changes from unimodal with maximum at the trap center in the diffusive regime to a distribution sharply 
peaked at the boundaries.}
 \label{FprimeHO}
\end{figure}

We introduce the quantity
\begin{equation}\label{sigmaHO}
 \sigma(\beta_{0})=\frac{d}{d(\ln\beta_{0})}\ln\left[l_{\text{pers}}(\beta_{0}\mathcal{Z})^{-1}\right],
\end{equation} 
which can be interpreted, analogously with the thermodynamic relation $\beta E=\partial\ln\mathcal{Z}^{-1}/\partial\ln\beta$, as the ratio of the internal energy [which in 
the 
overdamped limit is given solely by the average of the potential energy, $E=\langle U(x)\rangle$] to the effective thermal energy of the system, $k_{B}T_{0}$, i.e., 
$\sigma(\beta_{0})=E/k_{B}T_{0}$. Equation \eqref{sigmaHO} can be written in terms of the digamma function, $\Psi(x)=\left[\ln\Gamma(x)\right]^{\prime}$, as
\begin{equation}
 \sigma(\beta_{0})=\frac{1}{2\beta_{0}}\left[\Psi\left(\frac{1}{2}+\frac{1}{2\beta_{0}}\right)-\Psi\left(\frac{1}{2\beta_{0}}\right)\right].
\end{equation}
As is apparent from Fig. \ref{FprimeHO}, $\sigma(\beta_{0})$ is a sigmoidlike function of $\beta_{0}$. This characteristic make it suitable as a measure of the 
departure from the Boltzmann-Gibbs distribution. Certainly, in the limit of negligible persistence, i.e.,  $\beta_{0}\ll1$, one can substitute the Gamma functions that appear 
in Eq. \eqref{ZHO} by their Stirling's approximation to get $\sigma(\beta_{0})\approx 1/2$ as is shown in Fig. \ref{FprimeHO}. This result can be interpreted as the 
fulfillment of the equipartition theorem. In contrast, it can be shown that in the limit of large persistence, $\beta_{0}\gg1$, 
$\sigma(\beta_{0})$ saturates to the value 1, which characterizes the non-Boltzmann-Gibbs distribution for which the fulfillment of the equipartition theorem breaks down, we 
get $E=k_{B}T_{0}$. 

\subsubsection{Steeper power law trapping potentials}
Due to the fact that the harmonic potential \eqref{HarmonicPot} satisfies that $[U^{\prime}(x)]^{2}$ is proportional to 
$U(x)$ itself, the thermophoretic force [computed from Eq. \eqref{TempProfileParticular}] opposes the force derived from the trapping potential with the 
same dependence on the position, linear in this case, which leads to the stationary solution \eqref{SolStationaryHO}. It is clear that more subtle effects should appear
if steeper potentials are considered. We briefly point out some aspects for a power-law potential of the form  
\begin{equation}\label{PowLawPot}
 U(x)=\dfrac{\kappa_{n}}{2n}x^{2n},\quad n=1,2,\ldots,
\end{equation}
where $\kappa_{n}$ is a parameter with units of energy over [length]$^{2n}$ that indicates the strength of the potential.
Analogously with the parameter \eqref{beta0HO}, we introduce the parameter $\beta_{0,n}$ given by
\begin{equation}\label{beta0PowerLawPot}
 \beta_{0,n}=\kappa_{n}\mu\frac{v^{2(n-1)}}{\alpha^{2n-1}}=\left[\frac{l_{\text{pers}}}{x_{\text{max}}}\right]^{2n-1}
\end{equation}
that again measures the competence between the confinement length, $x_{\text{max}}=\left(v/\mu\kappa_{n}\right)^{1/(2n-1)}$, and the persistence length $l_{\text{pers}}$. A 
novel effect appears as consequence. Namely, it can be shown that for the potential as given in \eqref{PowLawPot}, new unstable equilibrium positions for 
the effective potential in Eq. \eqref{EffectivePotential} emerge for values of $\beta_{0,n}$ larger or equal to $[2(2n-1)]^{-(2n-1)}$ as long as $n>1$. The appearance of 
this unstable positions lead to \emph{multimodal} stationary distributions.

For the sake of clarity and for reasons that will be clear in the following section, where the symmetric double well potential is considered, we focus our analysis in the 
case $n=2$, the so-called \emph{quartic potential},
\begin{equation}\label{QuarticPot}
 U(x)=\kappa_{2}x^{4}/4,
\end{equation}
for which $\beta_{0,2}=\mu\kappa_{2}v^{2}/\alpha^{3}=\left(l_{\text{pers}}/x_{\text{max}}\right)^{3}$. The value of $\beta_{0,2}$ for which the unstable equilibrium positions 
of the effective potential start to appear is $6^{-3}$. This value also marks a qualitative difference of the effective potential, namely it has a vanishing slope at the 
boundaries. For larger values of $\beta_{0,2}$ the effective potential exhibits the mentioned 
unstable equilibrium (maxima) positions; however, the center of the trap is still the stable equilibrium position of the effective potential. Thus, the particles are ``pushed 
away'' from the unstable positions accumulating, on the one hand, 
at the boundaries, and, on the other, at the center of the trap leading to a multimodal probability distribution. This feature changes abruptly for $\beta_{0,2}>4^{-3}$, 
since in this regime the effective potential has its minima at the borders.  

\begin{figure}
 \includegraphics[width=\columnwidth]{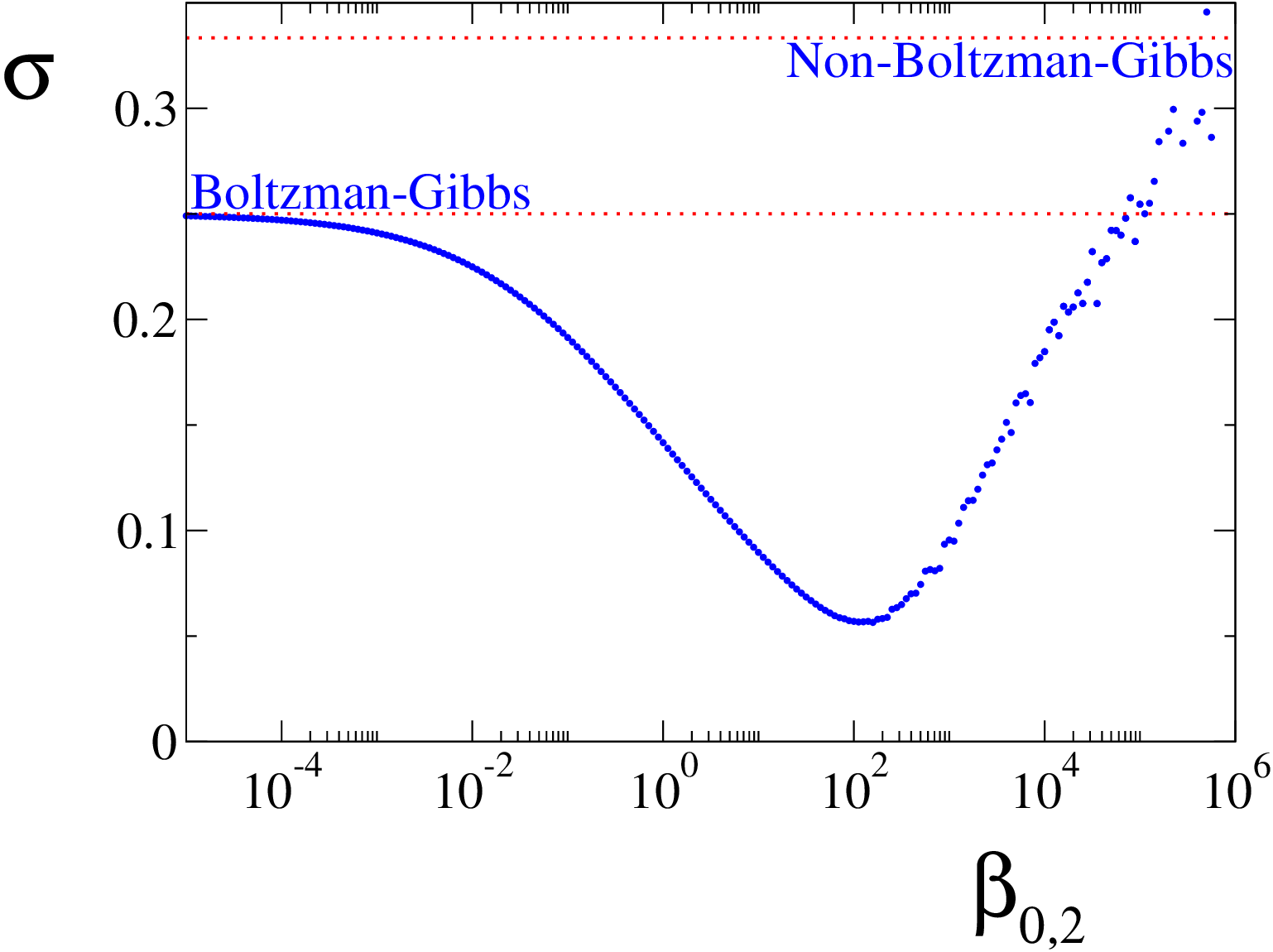}
 \caption{(Color online) $\sigma$ as function of $\beta_{0,2}$ is shown for the case of the quartic potential \eqref{QuarticPot}. It saturates in the persistent 
regime ($\beta_{0,2}\gg1$) at the value $\frac{1}{3}$, while in the diffusive regime ($\beta_{0,2}\ll1$) $\sigma$ characterizes the Boltzmann-Gibbs distribution with the 
value $\frac{1}{4}.$ 
}
 \label{SigmaQuarticPotential}
\end{figure}

In complete analogy with Eq. \eqref{sigmaHO}, we consider the quantity
\begin{equation}\label{sigmaQuartic}
 \sigma(\beta_{0,2})=\frac{d}{d(\ln\beta_{0,2})}\ln\left[l_{\text{pers}}(\beta_{0,2}\mathcal{Z})^{-1}\right]
\end{equation} 
as a measure of the departure from the stationary distribution of Boltzmann-Gibbs; however,
in contrast with the case of the harmonic potential, no analytical expression of the partition function for arbitrary $\beta_{0,2}$ exists. Notwithstanding this, it is 
straightforward to show that in the diffusive limit  $\sigma(\beta_{0,2})\rightarrow1/4$, while in the asymptotic limit, $\beta_{0,2}\rightarrow\infty$, we have that 
$\sigma(\beta_{0,2})\rightarrow1/3$ as is shown in the Fig. \ref{SigmaQuarticPotential}.

\subsection{The symmetric double-well potential}
A more interesting case corresponds to the one where the trapping potential has a richer energy landscape as occur with those that exhibit more than one stable state. Here, 
we 
report on the simplest case when the trapping potential has only two degenerate stable states. Thus, we focus our analysis in the effects of persistence of active particles 
trapped by the symmetric double-well potential,
\begin{equation}\label{DoubleWell}
U\left(x\right)=\Delta U \left[\frac{x^{4}}{L^4}-2\frac{x^{2}}{L^2}+1\right],
\end{equation}
leaving for a further analysis the effects of active motion on the asymmetric one as the one considered in Ref. \cite{LandauerPRA1975} by Landauer. In Eq. 
\eqref{DoubleWell} $\Delta U$ and $L$ are two positive parameters that 
characterize the external potential, the former one denotes the energy height of the barrier, while the last one corresponds to half the distance between the two stable 
states that correspond to the minima of the external potential \eqref{DoubleWell} located at $x=\pm L$, respectively.

\begin{figure}
\includegraphics[width=\columnwidth]{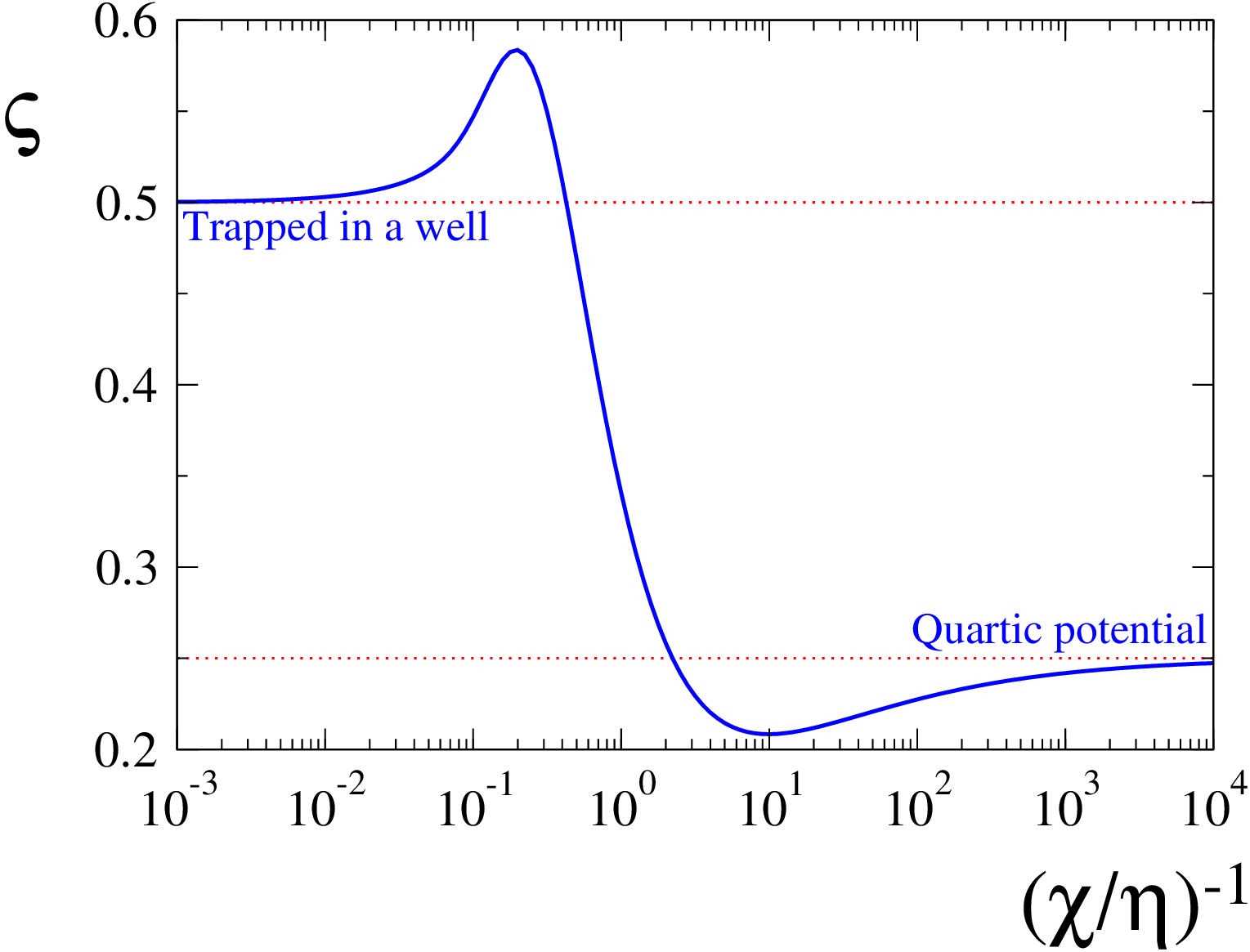}
\caption{(Color online) Susceptibility $\varsigma$ as in Eq. \eqref{SigmaDW-Boltzmann}, as function of the dimensionless persistence length $(\chi/\eta)^{-1}$. For 
$(\chi/\eta)^{-1}\ll1$ the particle is trapped in any of the potential wells, and in the opposite regime $(\chi/\eta)^{-1}\gg1$ the quartic dependence of the potential is the 
dominant part of the trapping.}
\label{FigSigmaBetaChizero}
\end{figure}

In addition to the characteristic lengths $l_{\text{pers}}$ and $L$, there is a third one,
\begin{equation}
 \mathcal{L}=\frac{8}{3\sqrt{3}} \Delta U\frac{\mu}{v},
\end{equation}
that gives an estimate of the average length the active particle travels from the center of the potential, when an energy $\Delta U$ is available to swim at a swimming 
force $v/\mu$. For convenience we introduce the following two independent parameters:
\begin{subequations}
 \begin{align}
  \chi&=\frac{\mathcal{L}}{L},\\
  \eta&=\frac{l_{\text{pers}}}{L}.
 \end{align}
\end{subequations}
Small values of $\chi$ refer to either a shallow energy barrier,  potential wells far apart, or both. Besides these, we introduce a third parameter,
\begin{equation}
 \beta_{\text{dw}}=4\frac{\mu\, \Delta U}{L^{2}\alpha}=\frac{3\sqrt{3}}{2}\chi\eta,
\end{equation}
which characterizes the departure from the equilibrium regime and allows the transition between the Boltzmann-Gibbs distribution and its corresponding stationary 
superstatistics distribtuion in the active (persistence) regime. Indeed, if the effects of persistence of active motion are negligible, as occurs in the diffusive regime, 
then we have that $\beta_{\text{dw}}\rightarrow0$ and the stationary distribution corresponds to the one of Boltzmann-Gibbs 
\begin{equation}
 P_{\text{B-G}}(x)=\mathcal{Z}^{-1}\exp\left\lbrace-\frac{3\sqrt{3}}{8}\frac{\chi}{\eta}\left(\frac{x^{4}}{L^{4}}-2\frac{x^{2}}{L^{2}}+1\right)\right\rbrace,
\end{equation}
which corresponds to the bimodal distribution that is symmetric at the stable positions of the external potential $\pm L$. In this limit ($\beta_{\text{dw}}\rightarrow0$) we 
are interested in the quantity analogous to the one given in Eq. \eqref{sigmaHO}, in this case given by
\begin{equation}\label{SigmaDW-Boltzmann}
 \varsigma(\chi/\eta)=\frac{d}{d[\ln(\chi/\eta)]}\ln(L\mathcal{Z}^{-1}),
\end{equation}
which is shown in Fig. \ref{FigSigmaBetaChizero}, as function of $(\chi/\eta)^{-1}$. 
For large values of $\chi/\eta$, i.e., $\Delta U \gg k_{B}T_{0}$, the system is reminiscent of a system of passive Brownian particles that get trapped at the minima of the 
potential and $\varsigma(\chi/\eta)\rightarrow\frac{1}{2}$. In the opposite regime $\Delta U\ll k_{B}T_{0}$, the effects of the energy barrier are negligible and the 
stationary distribution corresponds to that of a trapped passive Brownian particle diffusing in 
a quartic potential leading to $\varsigma(\chi/\eta)=\frac{1}{4}$. The transition between these limit values is not monotonous as is clear from the figure.

As has been discussed above, in the stationary regime, the nonequilibrium nature of trapped active motion can be mapped into a fictitious inhomogeneous thermal bath 
characterized by the temperature profile given by \eqref{TempProfileParticular}, where trapped passive Brownian particles diffuse. As was pointed out by Landauer 
\cite{LandauerPRA1975}, such effects due to inhomogeneity can change the relative stability of the system stable states. Interestingly, we show that active motion give rise to 
the appearance of metastable states if the persistence length overpasses a threshold value of the confinement length.

The spatial dependence of the temperature profile associated to the potential \eqref{DoubleWell} is given by 
\begin{equation}\label{DoubleWellTemp}
 \mathcal{T}(x)=T_{0}\left\{1-\beta_{\text{dw}}^{2}\frac{x^{2}}{l_{\text{pers}}^{2}}\left(\frac{x^{2}}{l_{\text{pers}}^{2}}\eta^{2}-1\right)^{2}\right\},
\end{equation}
which reduces to the homogeneous effective temperature $T_{0}$ as $\beta_{\text{dw}}\rightarrow0$.
The temperature profile \eqref{DoubleWellTemp} has a richer structure (see Fig. 
\ref{TempProfile}) in comparison with profiles associated with external potentials with only one stable state. By simple inspection it can be seen that the temperature 
profile 
\eqref{DoubleWellTemp} reaches its maximum value $T_{0}$, at the equilibrium positions of the external potential either stable or unstable, namely at $x=\pm L$ 
and at the origin $x=0$, respectively. Additionally, $\mathcal{T}(x)$ has two local minima at the positions $x=\pm L/\sqrt{3}$, where the local temperature acquires the value 
$T_{0}\left\{1-\chi^{2}\right\}$. 
\begin{figure}
\begin{overpic}[scale=.3175]{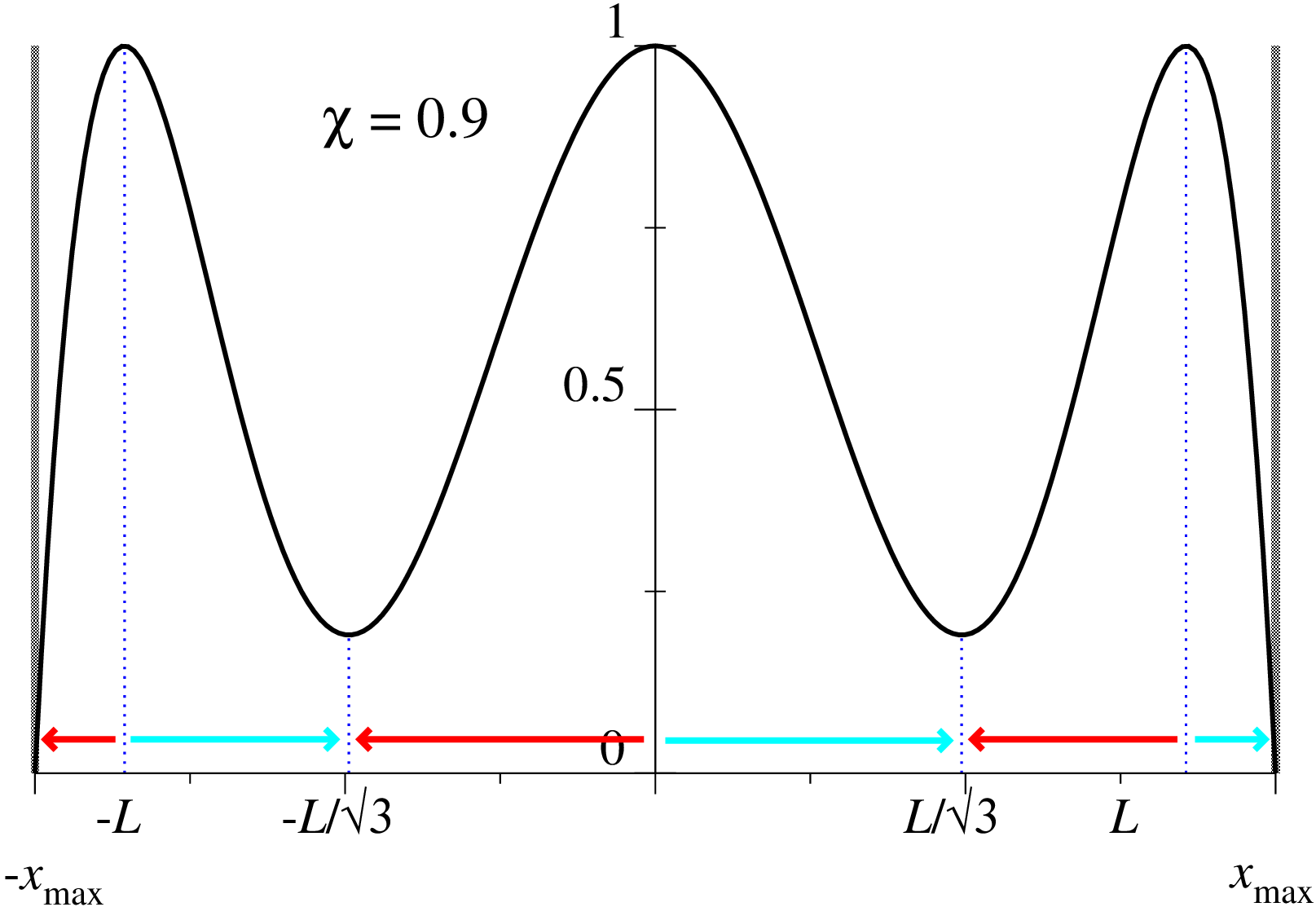}
\put(55,65){\large $\mathcal{T}(x)/T_{0}$}
\end{overpic} 
 \caption{The dimensionless effective temperature $\mathcal{T}(x)/T_{0}$ for the double well potential \eqref{DoubleWellTemp} with $\chi=0.9$. The direction of the 
thermophoretic force 
\eqref{DoubleWellThermalF} is shown by arrows along the horizontal axis.}
 \label{TempProfile}
\end{figure}

We focus our analysis to the case for which the inequality $\chi<1$ is valid, first, because this condition guaranties a positive definite local temperature and, second, it 
avoids the unnecessary difficulty of defining the flux in the regions of space where this conditions is not valid. In simple words, this 
condition assures that the particle is active enough to overcome the energy barrier [$v>(4/3)^{3/2}\, \mu\, \Delta U/L$] and makes it able 
to freely swim between the two boundary points $\pm x_{\text{max}}$, given by
\begin{multline}\label{xmax}
x_{\text{max}}= \frac{L}{\sqrt{3}}\left[\chi^{1/3}\left(1+\sqrt{1-\chi^{2}}\right)^{-1/3}+\right.\\
\left.\chi^{-1/3}\left(1+\sqrt{1-\chi^{2}}\right)^{1/3}\right],
\end{multline}
at which the net force on the particle vanishes. As $\chi\rightarrow1$ from below, the local effective temperature vanishes at $\pm L/\sqrt{3}$ as also does the net force on 
the particle.

The thermophoretic force induced by the effective local temperature \eqref{DoubleWellTemp} is explicitly given by the product of the swimming force $v/\mu$ times a factor 
that takes into account the inhomogeneity of the fictitious medium, namely
\begin{equation}\label{DoubleWellThermalF}
 2\frac{v}{\mu}\beta_{\text{dw}}\frac{x}{l_{\text{pers}}}\left(\frac{x^{2}}{l^{2}_{\text{pers}}}\eta^{2}-1\right)\left(3\frac{x^{2}}{l^{2}_{\text{pers}}}\eta^{2}-1\right),
\end{equation}
which, for finite $\beta_{\text{dw}}$, pushes the particles away from the positions of maximum temperature toward either the boundaries $\pm x_{\text{max}}$ or toward 
$\pm L/\sqrt{3}$, at which 
$\mathcal{T}(x)$ has local minima. Thus there is a competition between the thermophoretic force that tends to accumulate the particles at these positions (such a situation is 
pictorially shown in Fig. \ref{TempProfile} for a temperature profile characterized by $\chi=0.9$) and the force due to the external potential $U(x)$ that tends to accumulate 
the particles at $x=\pm L$. 
\begin{figure}
 \includegraphics[width=\columnwidth]{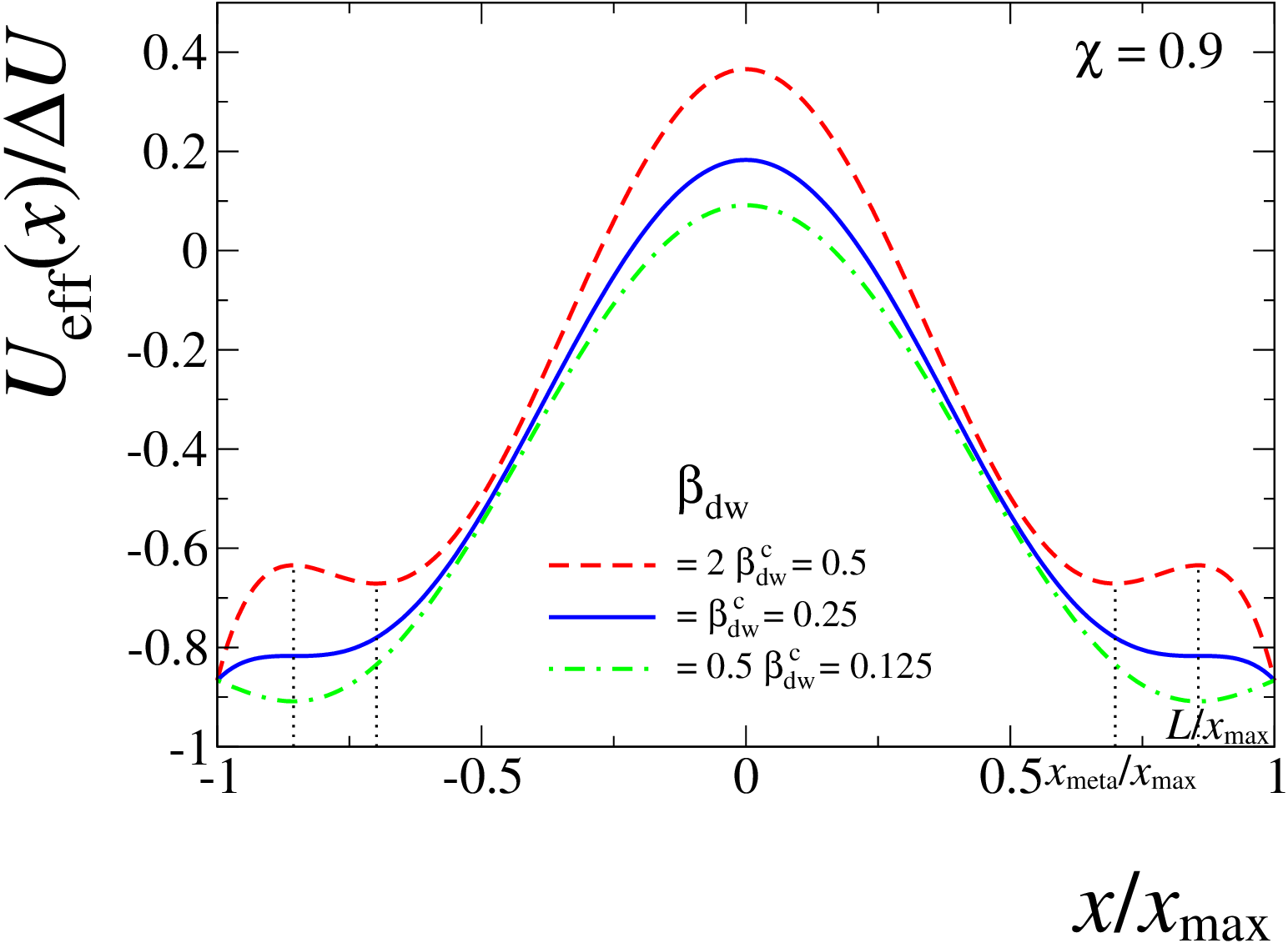}
 \caption{(Color online) The dimensionless effective potential, $U_{\text{eff}}(x)/\Delta U$, as function of the dimensionless position $x/x_{\text{max}}$ for $\chi=0.9$. The 
blue solid line corresponds to the effective potential at the critical value $\beta_{\text{dw}}^{\text{c}}=\frac{1}{4}$. The dashed-dotted-green line corresponds to 
the value below threshold $\frac{1}{2}\beta_{\text{dw}}^{\text{c}}=\frac{1}{8}$, for which the $U_{\text{eff}}(x)$ has global minima at the positions $\pm L$. The 
dashed-red line shows the behavior of $U_{\text{eff}}(x)$ for the value above threshold $2\beta_{\text{dw}}^{\text{c}}=\frac{1}{2}$. The emergence of two metastable states 
at the positions $x_{\text{meta}}$ given by Eq. \eqref{PositionsMeta} is marked with dotted lines.}
 \label{FigEffectiveU}
\end{figure}

An straightforward stability analysis of $U_{\text{eff}}(x)$ shows that for a given value of $\chi$, the landscape of the effective potential changes as $\beta_{\text{dw}}$ 
is varied across the threshold value 
\begin{equation}\label{CriticalBeta}
 \beta_{\text{dw}}^{\text{c}}=\frac{1}{4}.
\end{equation}
Indeed, for $\beta_{\text{dw}}<\beta_{\text{dw}}^{\text{c}}$, the positions $\pm L$ correspond to global minima of $U_{\text{eff}}(x)$ (dashed-dotted green line in Fig. 
\ref{FigEffectiveU}, with $\chi=0.9$), which coincides with the minima of $U(x)$. These positions, however, become local maxima for 
$\beta_{\text{dw}}>\beta_{\text{dw}}^{\text{c}}$, and $U_{\text{eff}(x)}$ acquires local minima at the positions $\pm x_{\text{max}}$ and at the positions $\pm 
x_{\text{meta}}$, which emerge 
as metastable sates (dashed red line in the same figure), where 
\begin{equation}\label{PositionsMeta}
 x_{\text{meta}}=\frac{L}{\sqrt{3}}\left[1+\frac{1}{2}\beta_{\text{dw}}^{-1}\right]^{1/2}.
\end{equation}
At the critical value \eqref{CriticalBeta} the positions $\pm L$  are inflexion points and $\pm x_{\text{max}}$ become the global 
minima of $U_{\text{eff}}(x)$ (blue-solid line in Fig. \ref{FigEffectiveU}). 

\begin{figure}
 \includegraphics[width=\columnwidth]{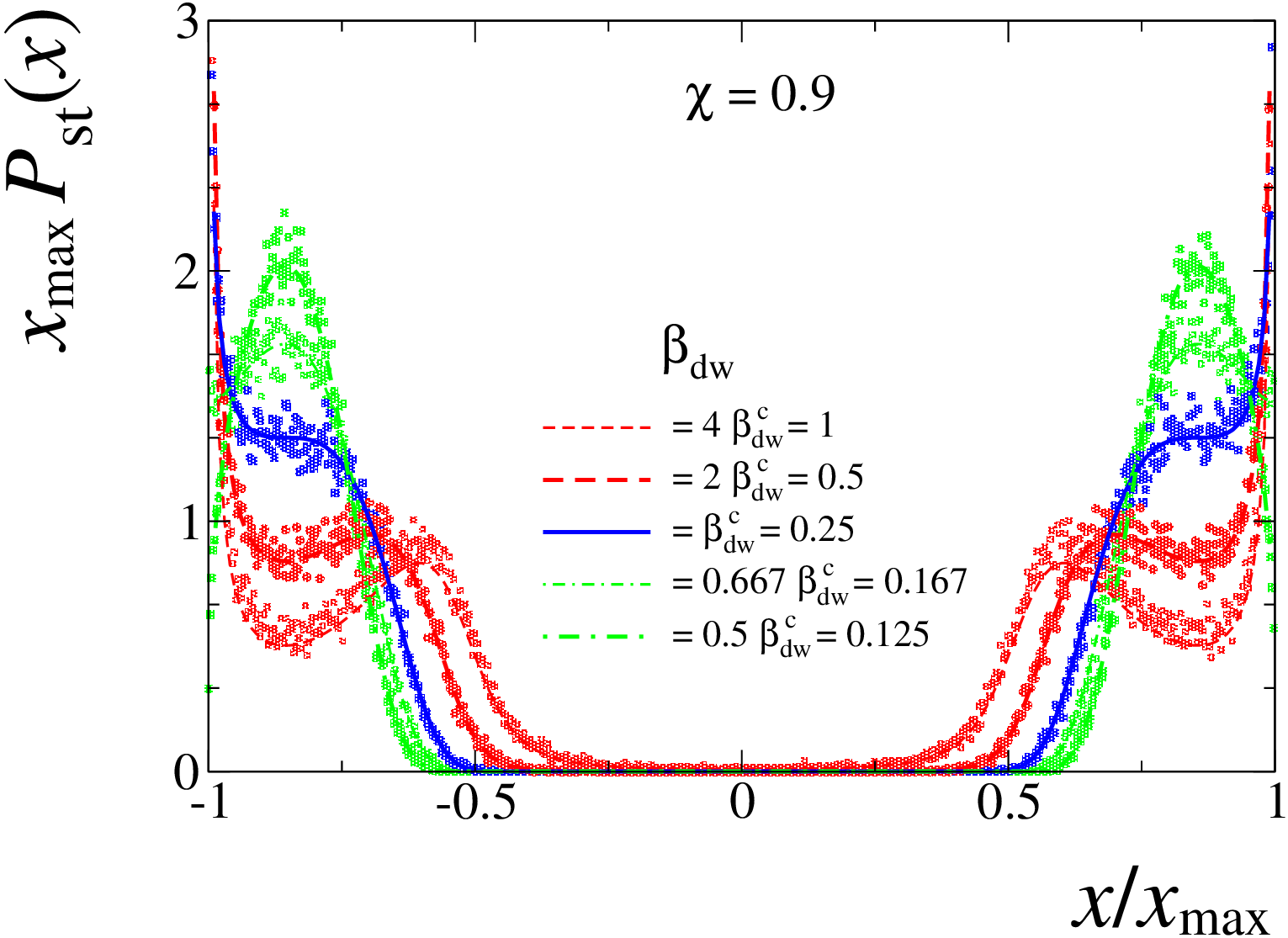}
 \caption{(Color online) The dimensionless stationary probability density function, $x_{\text{max}}P_{\text{st}}(x)$, as function of the dimensionless position 
$x/x_{\text{max}}$ for $\chi=0.9$ and different values of $\beta_{\text{dw}}$. The solid blue line corresponds to the threshold value 
$\beta_{\text{dw}}^{c}=\frac{1}{4}$. The thin- and thick-dashed green lines correspond to the values $\beta_{\text{dw}}=\frac{1}{6}$ and 
$\beta_{\text{dw}}=\frac{1}{8}$ below the threshold value, while the thin- and thick-dashed red lines, $\beta_{\text{dw}}=0.5$ and 
$\beta_{\text{dw}}=1$, respectively, correspond to the values of $\beta_{\text{dw}}$ above threshold. The symbols mark the probability 
density function calculated from the data obtained from numerical solution of Eq. \eqref{eq_langevin}.}
 \label{FigPDF_chi09}
\end{figure}

\begin{figure}
 \includegraphics[width=\columnwidth]{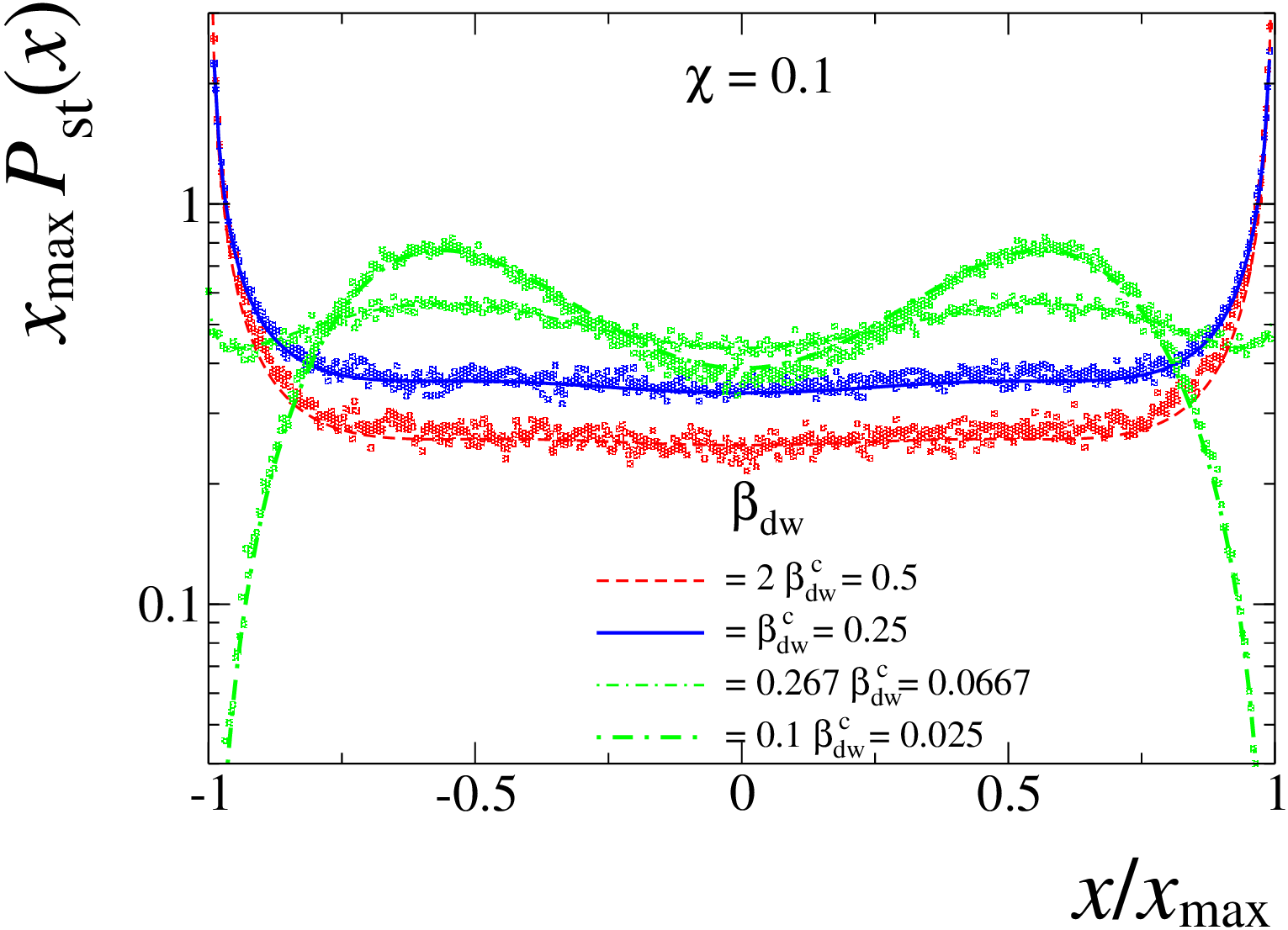}
 \caption{(Color online) The dimensionless stationary probability density function, $x_{\text{max}}P_{\text{st}}(x)$, as function of the dimensionless position 
$x/x_{\text{max}}$ for $\chi=0.1$ and different values of $\beta_{\text{dw}}$. The solid blue line corresponds to the threshold value $\beta_{\text{dw}}^{c}=\frac{1}{4}$.
The thin- and thick-dashed green lines correspond to the values  $\frac{4}{15}\beta_{\text{dw}}^{\text{c}}=\frac{1}{15}$ and 
$\frac{1}{10}\beta_{\text{dw}}^{\text{c}}=\frac{1}{40}$ below threshold, while $2\beta_{\text{dw}}^{\text{c}}=\frac{1}{2}$, given by the dashed red line corresponds to the 
value of $\beta_{\text{dw}}$ above $\beta_{\text{dw}}^{c}$. The symbols mark the probability density function 
calculated from the data obtained from numerical solution of Eq. \eqref{eq_langevin}.}
 \label{FigPDF_chi01}
\end{figure}

The dimensionless probability density function in the stationary state, $x_{\text{max}}P_{\text{st}}(x)$, as function of $x/x_{\text{max}}$ is shown for the values of 
$\chi=0.9$ and 0.1 in Figs. \ref{FigPDF_chi09} and \ref{FigPDF_chi01}, respectively. This is computed numerically from the formulas \eqref{EquilSol} and \eqref{DoubleWell} 
(shown in solid lines) since no closed analytical expression is available and it is compared with the corresponding distribution obtained from the numerical simulations 
(symbols) using the stationary Brownian dynamics as has been explained in Sec. \ref{section3}. In each figure, the mentioned distributions are shown for different values 
of the parameter $\beta_{\text{dw}}$. 
For the case $\chi=0.9$ (see Fig. \ref{FigPDF_chi09}), the thermophoretic force pushes the particles away from the center of the trap, and thus the probability density at the 
center is small. At the threshold value $\beta_{\text{dw}}^{\text{c}}$ (blue-solid line), the probability density grows monotonically 
from the center of the potential toward the boundaries, where the particles accumulate. For subcritical values, $\beta_{\text{dw}}<\beta_{\text{dw}}^{\text{c}}$, the effects 
of persistent motion are diminished and the particles accumulate at $x=\pm L$ (see dashed-dotted green lines in Fig. \ref{FigPDF_chi09} for $\beta_{\text{dw}}=\frac{1}{6}$ and 
$\beta_{\text{dw}}=\frac{1}{8}$), as would occur in the equilibrium case. As $\beta_{\text{dw}}$ is further decreased, the probability distribution of the particle positions 
tends to the symmetrically bimodal distribution, proportional to the Boltzmann-Gibbs factor $\exp\{-U(x)/k_{B}T_{0}\}$. In the regime above threshold, 
$\beta_{\text{dw}}>\frac{1}{4}$, the landscape of the effective potential changes and the stationary probability density shows two symmetrically pairs of peaks (see dashed red 
lines in Fig. \ref{FigPDF_chi09} for $\beta_{\text{dw}}=\frac{1}{2}$ and $\beta_{\text{dw}}=1$), i.e., by increasing of effects of persistent motion the modality of the 
distribution is enhanced, going from bimodal to four modes. Indeed, the most populated modes correspond to the pair located at the boundaries, while the other pair of modes 
are located at $\pm x_{\text{meta}}$. In the asymptotic limit $\beta_{\text{dw}}\rightarrow\infty$, the positions of the metastable states, $\pm x_{\text{meta}}$, coincide 
with the local minima of the temperature profile $x=\pm L/\sqrt{3}$, while the positions $\pm L$ become local minima. 

In Fig. \ref{FigPDF_chi01}, the stationary probability density function is shown for $\chi=0.1$, a value that corresponds to shallow energy barriers for which the 
particle can explore the whole space available between $-x_{\text{max}}$ and $x_{\text{max}}$ without being hindered by the barrier, as occurs in the previous case. 
At the threshold value $\beta_{\text{dw}}=\frac{1}{4}$ (blue-solid line), the probability density is almost flat at the center of the potential and it peaks at the boundaries 
$\pm x_{\text{max}}$. As the effect of persistent motion is diminished, i.e., for $\beta_{\text{dw}}<\frac{1}{4}$, the peaking of the probability density function at 
the boundaries is diminished gradually until particles start to accumulate at $\pm L$. In the diffusive limit, the region close to the boundaries is visited much less 
frequently (see dashed-dotted green lines in Fig. \ref{FigPDF_chi01} for $\beta_{\text{dw}}=\frac{1}{15}$ and $\beta_{\text{dw}}=\frac{1}{40})$. If $\beta_{\text{dw}}$ is 
further decreased as before, then the probability distribution of the particle positions tends to the equilibrium distribution characterized by the Boltzmann-Gibbs factor. In 
contrast to the cases for which $\chi\lesssim1$, the peaking of the probability distribution at $x_{\text{meta}}$ is subtle, while the peaking at the boundaries is 
conspicuous in the supercritical regime (see the dashed red line in Fig. \ref{FigPDF_chi01} for $\beta_{\text{dw}}=\frac{1}{2}$).

\begin{figure}
 \includegraphics[width=\columnwidth]{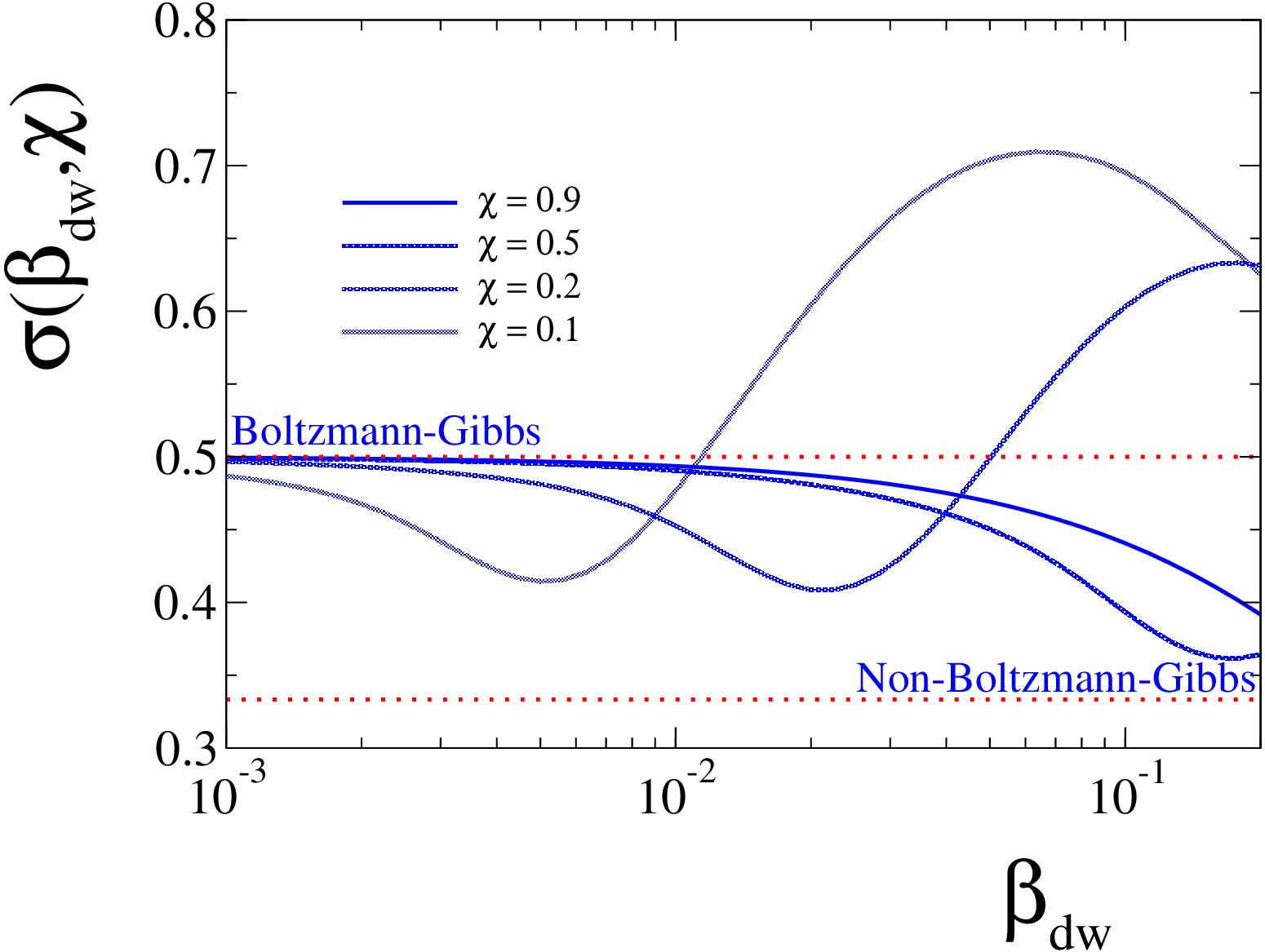}
 \caption{(Color online) $\sigma(\beta_{\text{dw}},\chi)$ as defined in Eq. \eqref{SigmadoubleWell} for the case of the double-well potential \eqref{DoubleWell}, is shown as a 
function of $\beta_{\text{dw}}$ for the values $\chi=0.1$, 0.2, 0.5, and 0.9. In the diffusive regime 
($\beta_{\text{dw}}\ll1$), $\sigma\simeq\frac{1}{2}$ characterizes the equilibrium stationary state that corresponds to the Boltzmann-Gibbs distribution. The persistence 
regime ($\beta_{\text{dw}}\gg1$), $\sigma\simeq\frac{1}{3}$, characterizes the non-Boltzmann-Gibbs stationary distribution. These characteristic limiting values have been 
obtained analytically from Eq. \eqref{SigmadoubleWell}.}
 \label{FigSigma_DW}
\end{figure}

We conclude the analysis of run-and-tumble particles trapped by the symmetric double-well potential \eqref{DoubleWell} by computing the analogous quantity introduced in the 
previous section [Eq. \eqref{sigmaHO}] given now by
\begin{equation}\label{SigmadoubleWell}
 \sigma(\beta_{\text{dw}},\chi)=\frac{d}{d(\ln\beta_{\text{dw}})}\ln\left\lbrace 
l_{\text{pers}}\widetilde{\mathcal{Z}}^{-1}(\beta_{\text{dw}},\chi)\right\rbrace,
\end{equation} 
where the explicit dependence on $\chi$ has been pointed out and $\widetilde{\mathcal{Z}}(\beta_{\text{dw}},\chi)$ denotes the rescaled partition function $\exp 
\left\{-S(\beta_{\text{dw}},\chi)\right\}\mathcal{Z}(\beta_{\text{dw}},\chi)$, where $S(\beta_{\text{dw}},\chi)$ is a shift that makes the argument of the exponential in 
the partition function \eqref{PartitionFunction} positive. In Fig. \ref{FigSigma_DW}, the numerically computed susceptibility $\sigma(\beta_{\text{dw}},\chi)$ as function of 
$\beta_{\text{dw}}$ is shown for $\chi=0.1$, 0.2, 0.5 and 0.9. A nonmonotonous dependence on $\beta_{\text{dw}}$ is observed. In the diffusive limit, for which the stationary 
distribution corresponds to that of Boltzmann-Gibbs, the susceptibility marks the value $\frac{1}{2}$ independently of the value of $\chi$. As the effects of persistence 
become conspicuous, the particle distribution peaks at the boundaries, and, as such, the details of trapping potential at the center potential can be neglected making the 
quartic potential the dominant part. Therefore it is expected that $\sigma(\beta_{\text{dw}},\chi)$ goes to $\frac{1}{3}$ as occurs in Fig. \ref{SigmaQuarticPotential}. The 
peaking of the distribution at the boundaries in this regime makes such a calculation computationally difficult.

\section{\label{section4}Final comments and concluding remarks}

As is well known, the coupling between the diffusion process of active particles of spatially-independent active-motion traits and the inhomogeneity induced by the external 
potential, makes explicit the nonequilibrium aspect of active motion revealed in the non-Boltzmann-Gibbs stationary distributions of the particle position. In this paper we 
have established a single-parameter mapping between these non-Boltzmann-Gibbs stationary distributions \eqref{EquilSol} of run-and-tumble particles constrained by an external 
potential and the corresponding ones of passive Brownian motion under the same trapping potential but in a \emph{fictitious} nonuniform-temperature medium. Such a mapping, 
given by the prescription \eqref{TempProfileParticular}, allows a simple interpretation of the intrinsic nonequilibrium aspects of active matter marked by the stationary 
non-Boltzmann-Gibbs distributions, namely it brings to mind a passive Brownian particle diffusing in a fictitious medium at \emph{local equilibrium}, a concept that extends 
to the nonequilibrium realm some fundamental thermodynamics quantities. 

The single parameter that characterizes the mapping corresponds to the ratio of the potential-dependent confinement length and the persistence length 
$l_{\text{pers}}$. The homogeneity of the fictitious media is recovered in the diffusive limit, i.e., in the limit when the persistence length is much smaller than 
the confinement length, which leads to the equilibrium distributions of Boltzmann-Gibbs as the stationary solutions and brings back the concept of effective 
temperature. 
In the persistence regime, when the persistence length is larger or of the order of the confinement length, the stationary distributions can be understood as superstatistics 
distributions. The particular superstatistics distributions called $q$-Gaussians appear in the case when the trapping potential corresponds the harmonic 
one (see Sec. \ref{SubSecHarmonicPotential}). 

More specifically, we have considered the simplest run-and-tumble particles trapped in an external potential, i.e., particles that swim at constant speed $v$ and tumble at a 
constant rate $\alpha$, as a nonequilibrium analog of the simplest system in equilibrium thermodyanmics, the trapped ideal gas. We have conveyed that the nonequilibrium 
feature corresponds to a specific inhomogeneity of a fictitious thermal bath whose temperature profile has the spatial dependence given by Eq. \eqref{TempProfileParticular}, 
where the parameters that characterize self-propulsion ($v$) and active fluctuations ($\alpha$) explicitly appear. A measure of it has been given by the susceptibilities 
$\sigma$'s, introduced in Eqs. \eqref{sigmaHO}, \eqref{sigmaQuartic}, \eqref{SigmaDW-Boltzmann}, and \eqref{SigmadoubleWell}. In addition, we find an explicit instance of the 
mechanism behind superstatistics \cite{BeckPRE2005}.  

We want to point out that the fluctuations in the swimming speed (not considered in the present analysis) causes a significant change in the nature of the stationary 
distributions. For instance, in the case of the so-called active Ornstein-Uhlenbeck model \cite{SzamelPRE2014} of active motion, where the swimming speed fluctuates according 
to an Ornstein-Uhlenbeck process, the stationary distribution of the particle position trapped by the harmonic potential is Gaussian, as in the equilibrium case but with an 
effective temperature. In such a case, active motion does not change the stability positions of the external trapping potential in contrasts with models that maintain the 
swimming speed constant. In a 
similar fashion as the analysis presented in this paper, it has been revealed that active Ornstein-Uhlenbeck motion, can be mapped into underdamped passive Brownian motion 
with a space dependent friction term, which can be understood as the coupling of the particle motion with a fictitious inhomogeneous medium that causes a local friction term 
\cite{CapriniJStatMech2018}. 

The extension of the present analysis to higher dimensions is not straightforward, indeed, neither the active Brownian nor the run-and-tumble model 
of active motion have explicit analytical solutions for arbitrary trapping potentials, thus leaving the determination of the mapping between the trapped active motion to 
passive Brownian motion in an inhomogeneous thermal bath as an open problem. On the other hand, the existence of the homogeneous effective temperature in the diffusive limit 
of two-dimensional active motion has been discussed in Ref.~\cite{Solon2015EPJST2015} which, as expected, is related with the equilibrium solution of Boltzmann-Gibbs in the 
zero-current stationary state. The existence of such homogeneous temperature has been shown to exists in a two-dimensional trapped system, namely for active 
Ornstein-Uhlenbeck particles trapped by a harmonic potential \cite{DasNJP2018}. Further, the analysis presented in this paper can be generalized to the case of 
one-dimensional run-and-tumble particles diffusing within a finite interval with reflecting boundary conditions (hard walls) at the borders \cite{MalakarJStatMech2018}, for 
which the high accumulation of particles at the boundaries can be mapped to passive Brownian motion in an inhomogeneous thermal bath under the same boundary 
conditions. Finally, it would be of broad interest to find a generalization of the Boltzmann-Gibbs entropy functional from which \eqref{GralSolution} is derived via the 
maximization of entropy principle, as has been shown from many non-Bolztmann-Gibbs distributions that occur in complex systems \cite{SchwammleEPJB2007}.

\begin{acknowledgments}
F.J.S kindly acknowledges support from Grant No. UNAM-DGAPA-PAPIIT-IN114717 and A.V.A. acknowledges support from Grant No. UNAM-DGAPA-PAPIIT-IA104917.
\end{acknowledgments}

\end{document}